 \documentclass[final,3p,times]{elsarticle}

\usepackage{amssymb}
\usepackage{amsmath}
\usepackage{color}

\makeatletter

\newcommand{\Rmnum}[1]{\expandafter\@slowromancap\romannumeral #1@}
    \def\ps@pprintTitle{%
      \let\@oddhead\@empty
      \let\@evenhead\@empty
      \let\@oddfoot\@empty
      \let\@evenfoot\@oddfoot
    }
\makeatletter

\begin{document}

\begin{frontmatter}



\title{Decomposition method for block-tridiagonal matrix systems}


\author[]{P.~A.~Belov\corref{cor1}}%
\ead{pavelbelov@gmail.com}
\author[]{E.~R.~Nugumanov\fnref{fn1}}%
\author[]{S.~L.~Yakovlev}%

\address{Department of Computational Physics, Saint-Petersburg State University, Ulyanovskaya 1, 198504 Saint-Petersburg, Russia}
\cortext[cor1]{Corresponding author}
\fntext[fn1]{Now at BashNIPIneft LLC, 86/1, Lenin Street, 450006 Ufa, Russia}
%

\begin{abstract}
The decomposition method which makes the parallel solution of the block-tridiagonal matrix systems possible is presented.
The performance of the method is analytically estimated based on the number of elementary multiplicative operations for its parallel and serial parts.
The computational speedup with respect to the conventional sequential Thomas algorithm is assessed for various types of the application of the method.
It is observed that the maximum of the analytical speedup for a given number of blocks on the diagonal is achieved at some finite number of parallel processors.
The values of the parameters required to reach the maximum computational speedup are obtained.
The benchmark calculations show a good agreement of analytical estimations of the computational speedup and practically achieved results.
The application of the method is illustrated by employing the decomposition method to the matrix system originated from a boundary value problem for the two-dimensional integro-differential Faddeev equations.
The block-tridiagonal structure of the matrix arises from the proper discretization scheme including the finite-differences over the first coordinate and spline approximation over the second one.
The application of the decomposition method for parallelization of solving the matrix system reduces the overall time of calculation up to 10 times.
\end{abstract}

\begin{keyword}
block-tridiagonal matrix \sep decomposition method \sep Thomas algorithm \sep parallel solution \sep computational speedup \sep three-body systems \sep Faddeev equations


\end{keyword}

\end{frontmatter}


%




\section{Introduction}

The finite-difference discretization scheme for the multidimensional second order partial differential equations (PDEs) leads to a system with the block-tridiagonal matrix.
A precision of the numerical solution of the PDE depends on the number of knots over each dimension that usually approches thousands or even tens of thousands.
In this case, each block of the matrix becomes large and sparse, but the block-tridiagonal structure is kept.
A combination of the finite-difference method over one coordinate and the basis set methods~\cite{nr} over other coordinates generally
produces block-tridiagonal matrix with dense blocks.
The intermediate case of the basis set with local support, namely splines~\cite{Ahlberg}, results in the band blocks with bandwidth comparable with
the number of bands appeared from the finite-difference method.

An adaptation of
the Gauss elimination~\cite{Isaacson} to the block-tridiagonal systems is known as
the Thomas algorithm (TA)~\cite{ThomasEng} which is also called as the matrix sweeping algorithm~\cite{ThomasRus}.
In this algorithm, the idea of Gauss elimination is applied to the blocks themselves.
A stable and time proved realization of another Gauss elimination based techinque, LU decomposition, for general matrices is available in packages like LAPACK~\cite{Lapack350}.
The direct application of this realization of the LU decomposition is often not feasible because of the large size of the matrix.
As a result, the TA and special variants of LU decomposition are appropriate techniques for such large problems.

The TA is robust and quite fast, but serial and thus hardly parallelized.
The only available parallelization is at the level of operations with matrices.
This reason does not allow one to use the algorithm at the modern computational facilities efficiently.
In order to enable parallel and much faster solution, we have developed the decomposition method (DM) for the block-tridiagonal matrix systems.
The idea of the method consists in rearranging the initial matrix into equivalent one, namely with the ``arrowhead'' structure, which allows a parallel solution.
The matrix is logically reduced to some new diagonal independent blocks, sparse auxiliary ones along the right and bottom sides, and a coupling supplementary block.
The solution includes parallel inversion of the independent blocks and solving the supplementary problem. 
%
The DM includes using the TA for dealing with the new blocks of the initial matrix,
but the parallel structure and possible recursivity of the method lead to remarkable growth of the performance.

The speedup of the DM depends on the size of the supplementary matrix system.
By default, this system is solved sequentially by the TA. If its size is relatively small, then
the DM gives linear growth of the performance with increase of the number of computing units (processors).
As the nonparallelized part of the method steadily enlarges, the linear growth is slowed down.
The speedup with respect to the TA achieves
its maximum at some number of computing units and then decreases.
This issue can specifically be overcome by applying the DM recursively to each independent block and to the supplementary problem with the coupling matrix.
In this case, the maximum computational speedup with respect to the TA can be increased by several times in comparison to the nonrecursive application of the DM.

The concept
of rearranging the initial matrix into ``arrowhead'' form has already been proposed for ordinary tridiagonal matrices.
The method comes from the domain decomposition,
where the idea to divide a large problem into small ones which can be solved independently was introduced, see Ref.~\cite{Bjorstad} and references therein.
In Ref.~\cite{Ortega} this idea is illustrated by the example of tridiagonal matrix obtained from the finite-difference discretization of the
one-dimensional Laplace operator.
It is stated there that the matrix of the supplementary problem in this method is the smallest one among the similar block methods under consideration. Therefore, the method is supposed to be the most efficient.
Moreover, it is close to the cyclic reduction method~\cite{Stone} and, as it is shown in Ref.~\cite{Mehrmann}, is asymptotically equivalent to that method.
The DM is closely related to the so-called divide-and-conquer method implemented in ScaLAPACK for banded matrices~\cite{Cleary},
but designed directly to the block-tridiagonal matrix systems.

In this paper, we describe the DM for the block-tridiagonal matrix system in detail.
We show how one should solve the obtained
new rearranged system and provide the algorithm for it.
We underline in the text that the steps of the proposed algorithm can be executed in parallel.
The analysis of the number of multiplicative operations for the DM is given.
We analytically estimate the ratio of the multiplicative operations for the TA to the same quantity for different cases of application of the DM.
This ratio is directly related to the computational speedup and thus we estimate the performance of the DM with respect to the TA.
As a validation of the analytical results, we performed tests of the DM at the parallel supercomputing facility
which confirmed our estimations.

The block-tridiagonal matrix systems arise in many applied mathematics and physics problems.
We are unable to enumerate here all possible applications,
but briefly mention a few ones related to the quantum few-body systems.
The three-body scattering problem in configuration space has been firstly reduced to the boundary value problem in Ref.~\cite{M}
where the numerical technique for the solution has been also described.
This pioneering work gave rise to a class of papers inheriting the common approach for the numerical solution of the similar problems~\cite{M1986,Chen,Motovilov,BY}.
In the present paper, as an example of the application of the DM, we describe the computational scheme for solution of
the $s$-wave Faddeev integro-differential equations
and compare its performance with the TA or the DM.

The paper is organized as follows.
In Section~2, we describe in detail the DM and provide the algorithm for its implementation.
We give a brief description of the conventional TA in Section~3.
The analytical estimation of the number of multiplicative operations of the DM is presented in Section~4.
In Section~5, the computational speedup of the DM with respect to the TA
for various types of applications is given.
The desired parameters of the computational system for reaching the maximum speedup are also provided.
The validation of the analytical results is described in Section~6.
In Section~7, we describe a possible application of the DM for solving the $s$-wave Faddeev integro-differential equation
and give achieved time reduction.
Section~8 summarizes the article by providing the conclusion. 

\section{Decomposition method}

\begin{figure}
\begin{center}
\begin{tabular}{c}
\begin{minipage}{0.8\linewidth}
\includegraphics[width=1.0\textwidth, angle=0.0]{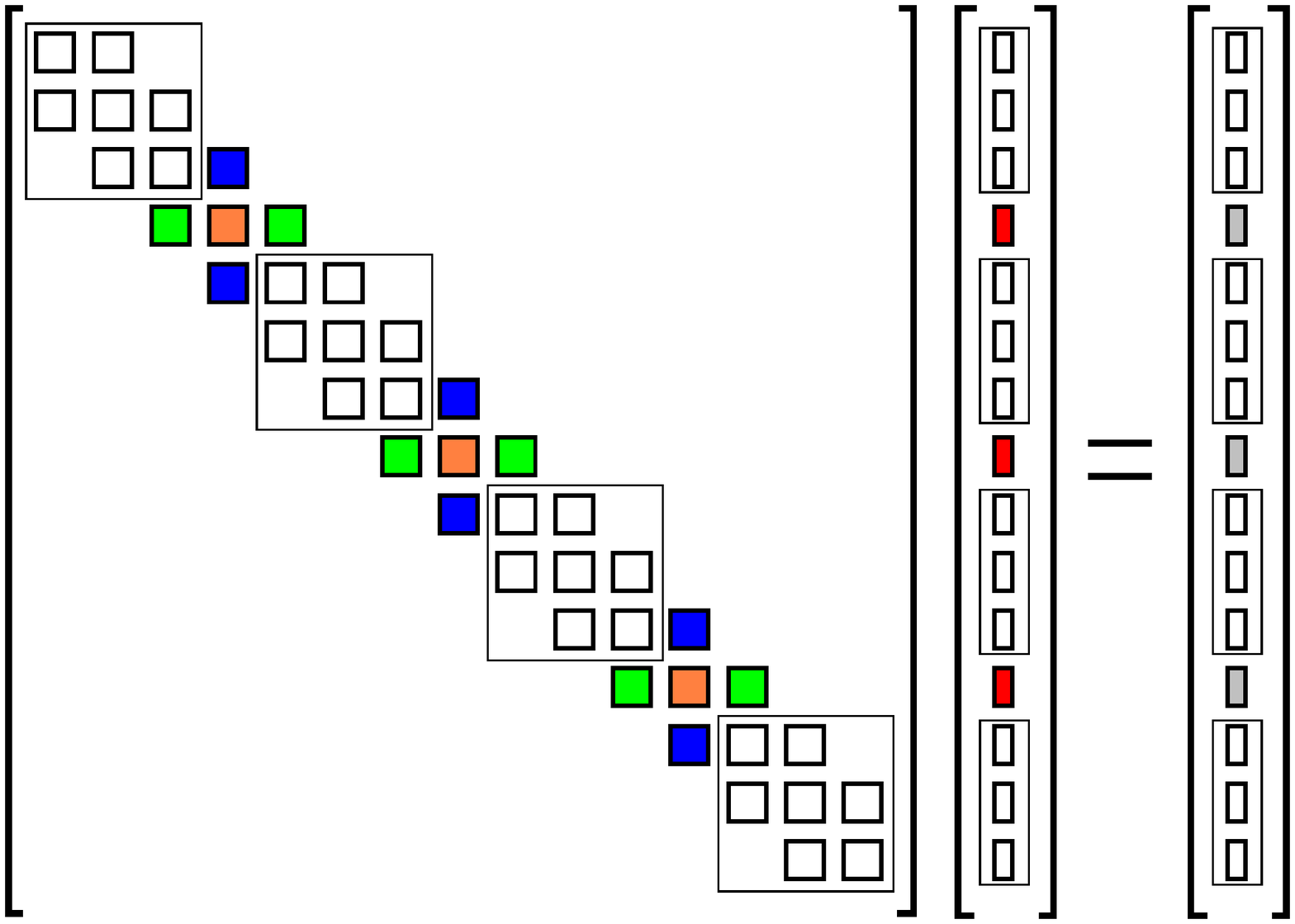}
\end{minipage} \\
\begin{minipage}{0.8\linewidth}
\includegraphics[width=1.0\textwidth, angle=0.0]{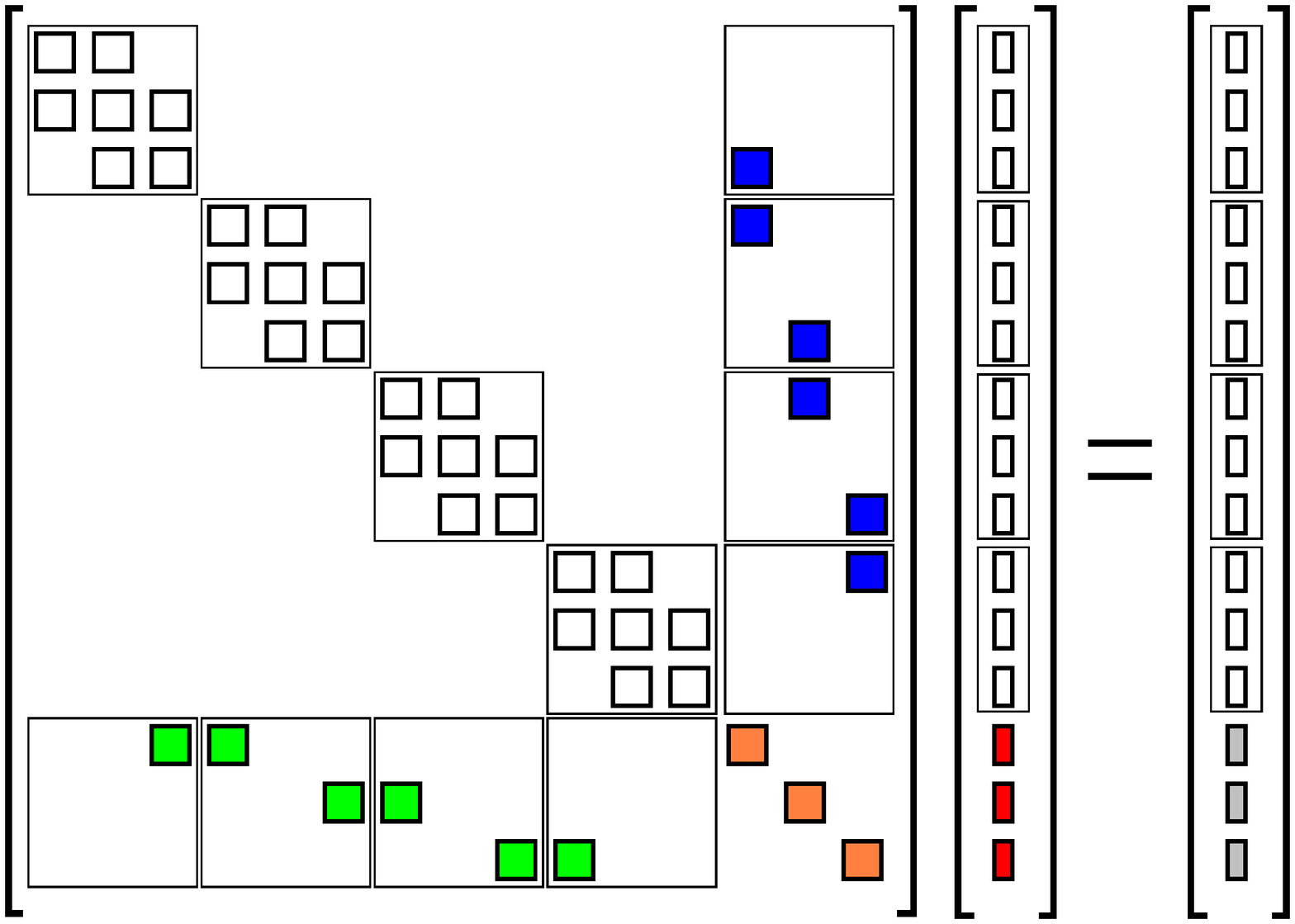}
\end{minipage}
\end{tabular}
\caption{\label{fig20}A graphical scheme of the rearrangement of the matrix system by the decomposition method.
Top panel: the initial matrix system with the colored separation blocks.
Bottom panel: the obtained rearranged system in the ``arrowhead'' form which can be solved using parallel calculations.
The nonzero blocks and vectors are denoted by thick lines, all other elements are trivial.
The new logical blocks and corresponding vectors at each panel are denoted by thin lines.}
\end{center}
\end{figure}

The block-tridiagonal matrix system under consideration is described by the equation
\begin{equation}
\label{eq01}
{A}_{i}X_{i-1}+{C}_{i}X_{i}+{B}_{i}X_{i+1}=F_{i}, \quad {A}_{1}={B}_{N}=0,
\end{equation}
where ${A}_{i}$, ${B}_{i}$, ${C}_{i}$, $i=1,\ldots,N$ are the blocks of the matrix
and $F_{i}$ are the blocks of the right-hand side (RHS) supervector ${\vec{F}}$.
The unknown supervector ${\vec{X}}$ is composed of blocks $X_{i}$.
The size of each block of the matrix is $n\times n$, whereas the size of each block of the supervectors is $n\times l$, where $l \ge 1$ is the number of columns.
The total size of the block-tridiagonal matrix is $(nN)\times (nN)$.

\begin{figure}[t!hp]
\centerline{\includegraphics[width=0.8\linewidth, angle=0.0]{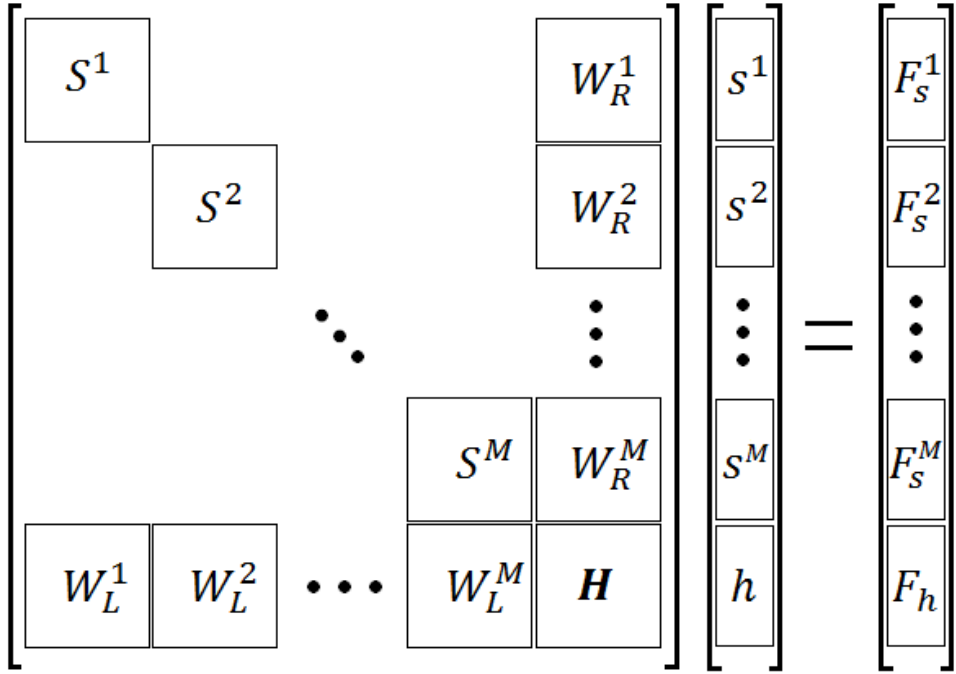}}
\caption{The conventional notation of the new logical blocks and corresponding vectors of the rearranged ``arrowhead'' matrix system.}
\label{fig021}
\end{figure}

The idea of the DM is presented in Fig.~\ref{fig20}.
The initial tridiagonal system~(\ref{eq01}) is rearranged into the equivalent, ``arrowhead'' form
which allows the parallel solving.
The rearrangement is performed in the following way:
\begin{itemize}
  \item One chooses a number of subsystems $M$ on which the initial matrix is divided. They are shown in Fig.~\ref{fig20} (top panel) by thin squares.
  \item $M-1$ separation
block-rows
of the whole system are arbitrary marked in rows from the second to the first before last. The block-columns corresponding to the diagonal block of each separation block-row are also marked.
  \item The marked separation block-rows together with the correspondent block-rows of the supervector in RHS are shifted to the bottom of the system by virtue of block-row interchanges. This procedure does not affect the elements of the unknown supervectors ${\vec{X}}$.
  \item The remained marked separation block-columns are shifted to the right part of the matrix. This movement affects the structure of the supervector $\vec{X}$ in such a way that the separation blocks of this vector are logically placed sequentially in the bottom, see Fig.~\ref{fig20} (bottom panel).
\end{itemize}
As a result, the initial system is logically rearranged into ``arrowhead'' form.
%
This new structure of the matrix system can be represented using the $2 \times 2$ block-matrix
which gives one a better insight into the way of solution.
The designed system can be written as
\begin{equation}
\label{e220}
\begin{pmatrix}
\mathbf{S} & \mathbf{W}_{R} \\
\mathbf{W}_{L} & \mathbf{H} \\
\end{pmatrix}
\begin{pmatrix}
s\\
h\\
\end{pmatrix}
=
\begin{pmatrix}
F_{s}\\
F_{h}\\
\end{pmatrix}.
\end{equation}
Here, the unknown solution $h$ corresponds to the moved part of the full solution.
The ``arrowhead'' form provides names for the matrix elements of Eq.~(\ref{e220}):
$\mathbf{S}$ comes from a ``shaft'', $\mathbf{H}$ is the ``head'' of an arrow, $\mathbf{W}_{R,L}$ are the right and left ``wings'' of the arrowhead.
This notation is shown in Fig.~\ref{fig021}.
The square
superblock $\mathbf{S}$ consists of the new independent blocks at the diagonal
$$
\mathbf{S}=\mbox{diag}\left\{ S^{1},\ldots,S^{M}\right\}.
$$
The matrix element $\mathbf{H}$ is the bottom right coupling superblock.
This element couples the independent blocks $S^{k}$ and consequently independent parts of the solution together to construct the complete one $\vec{X}$.
Other lateral superblocks $\mathbf{W}_{R}$, $\mathbf{W}_{L}$ present additional blocks of the matrix.
The solution of the system~(\ref{e220}) is given by the relations
\begin{equation}
\label{e230}
\left\{
\begin{array}{lcl}
s &=& \mathbf{S}^{-1} F_{s} - \mathbf{S}^{-1} \mathbf{W}_{R} h \\
h &=& \left( \mathbf{H} - \mathbf{W}_{L} \mathbf{S}^{-1} \mathbf{W}_{R} \right)^{-1} \left( F_{h} - \mathbf{W}_{L} \mathbf{S}^{-1} F_{s} \right)
\end{array}
\right. .
\end{equation}
The features of the numerical solving the system~(\ref{e230}) which affect the performance of the DM are following:
\begin{itemize}
  \item Due to the structure of obtained superblocks, the inversion of $\mathbf{S}$ is reduced to independent inversions of the diagonal blocks $S^{k}$, $k = 1,\ldots,M$ corresponding to each subsystem. These inversions can be performed in parallel.
  \item Since one has to calculate the
products
$\mathbf{S}^{-1} F_{s}$ and $\mathbf{S}^{-1} \mathbf{W}_{R}$ in Eq.~(\ref{e230}), 
it suffices to solve the equations
\begin{equation}
\label{e231}
\left\{
\begin{array}{lcl}
S^{k} Z^{k} = W^{k}_{R} \\
S^{k} z^{k} = F^{k}_{s}
\end{array}
\right. \quad k=1,\ldots, M,
\end{equation}
and is not needed to calculate inverse matrices $\left[ S^{k} \right]^{-1}$, $k = 1,\ldots,M$ explicitly.
%
%
  \item Once the solutions $Z^{k}$ and $z^{k}$ are obtained,
the matrix $\mathbf{H} - \mathbf{W}_{L} \mathbf{S}^{-1} \mathbf{W}_{R}$ and
the vector $F_{h} - \mathbf{W}_{L} \mathbf{S}^{-1} F_{s}$
are constructed
as
\begin{equation}
\label{e2311}
\begin{array}{lcl}
\mathbf{H} - \sum_{k=1}^{M} W^{k}_{L} Z^{k} \\
F_{h} - \sum_{k=1}^{M} W^{k}_{L} z^{k}.
\end{array}
\end{equation}
%
%
%
The fact that only two blocks of the matrices $W^{k}_{L}$ and $W^{k}_{R}$ are not trivial
drastically reduces the number of matrix operations in Eq.~(\ref{e2311}).
%
  \item The second line in Eq.~(\ref{e230}) can be treated as the supplementary matrix equation
\begin{equation}
\label{e232}
\left( \mathbf{H} - \mathbf{W}_{L} \mathbf{S}^{-1} \mathbf{W}_{R} \right) h = \left( F_{h} - \mathbf{W}_{L} \mathbf{S}^{-1} F_{s} \right).
\end{equation}
with the block-tridiagonal matrix.
The block-size of this system equals to $M-1$
and can be chosen much smaller than the size of the initial system (\ref{eq01}).
%
  \item Once the solution $h=(h^{1},\ldots,h^{M-1})^{T}$ of the supplementary matrix equation~(\ref{e232}) is obtained,
the remaining part $s$ of the complete solution $\vec{X}$ is calculated
independently for each subsystem as
\begin{equation}
\label{endSolution}
s^{k}=z^{k}-Z^{k}h^{k-1}-Z^{k}h^{k}, k=1,\ldots,M,
\end{equation}
where $h^{0}=h^{M}=0$.
\end{itemize}
As a result, the initial large matrix system is reduced to a set of independent small subsystems,
coupled to the initial one by the coupling matrix $\mathbf{H}$.
The number of subsystems $M$ should be chosen by a user in order to achieve the maximum performance of the DM.

To solve the independent subsystems in Eq.~(\ref{e231}) and the supplementary matrix equation~(\ref{e232})
one can apply any appropriate technique.
In our paper, for this purpose, we employ the TA as well as the DM itself.
The DM calls for the mentioned steps of the described algorithm lead us to the recursive application of the method.
This recursion allows one
to considerably improve overall performance of the method because it makes the previously serial part of the algorithm to be parallel.
In the following sections, we discuss in detail the application of the TA and the DM itself.

\section{The Thomas algorithm}

The conventional method for the direct solution of the block-tridiagonal matrix systems is the Thomas algorithm (TA)
which is also known as the matrix sweeping algorithm~\cite{ThomasEng,ThomasRus}.
The algorithm is based on the Gauss elimination technique applied to the block-tridiagonal structure of the matrix.
It includes two stages: the forward one and the backward one.
For the matrix equation~(\ref{eq01}),
the forward stage consists in the reduction of the initial matrix to the upper block-triangular matrix by sequential calculation of the auxiliary blocks
\begin{equation*}
\label{e200}
\left\{
\begin{array}{lcl}
\mathcal{A}_{1} &=& \mathcal{G}_{1}{B}_{1}\\
\mathcal{A}_{i} &=& \mathcal{G}_{i}{B}_{i},\quad i=2,...,N-1\\
\end{array}
\right.
\end{equation*}
and
\begin{equation*}
\label{e201}
\left\{
\begin{array}{lcl}
\mathcal{B}_{1} &=& \mathcal{G}_{1}{F}_{1}\\
\mathcal{B}_{i} &=& \mathcal{G}_{i} ({F}_{i}-{A}_{i}\mathcal{B}_{i-1}),\quad i=2,...,N,\\
\end{array}
\right.
\end{equation*}
where $\mathcal{G}_{1}={C}_{1}^{-1}$ and $\mathcal{G}_{i}=\left({C}_{i}-{A}_{i}\mathcal{A}_{i-1}\right)^{-1}$.
The backward stage sequentially yields the rows of the solution in the backward order using the calculated in the previous stage auxiliary blocks:
\begin{equation*}
\label{e210}
\left\{
\begin{array}{lcl}
X_{N} &=& \mathcal{B}_{N} \\
X_{i} &=& \mathcal{B}_{i} - \mathcal{A}_{i} X_{i+1},\quad i=N-1,\ldots,1.\\
\end{array}
\right.
\end{equation*}
As a result, the solution $\vec{X}$ of the matrix system~(\ref{eq01}) is obtained.

It is worth noting that the TA is serial:
one can obtain the next auxiliary block or next block of the solution vector only from the previous element.
Therefore, there is no possibility to make the TA to be parallel
except to execute the matrix operations parallel themselves at each step.

\section{Number of multiplicative operations}

The analytical estimation of the computational speedup is performed using the information on the number of multiplicative operations of the DM.
The additive operations are not taken into account because they are much less time consuming.
As a
benchmark
for estimation of the computational speedup, we consider the TA for the block-tridiagonal system.
Since the TA is sequential and easily implemented, it seems to be very convenient for comparison of the speedup.

According to type of the block-operations of the described algorithms,
for estimation of the computational speedup with respect to TA we have to take into account
the computational costs for the matrix operations: multiplication and inversion.
These matrix operations are already realized in linear algebra packages like BLAS and LAPACK.
The algorithms for them are well known~\cite{Gulin} and the computational costs have been estimated~\cite{Linpack}.

The number of multiplicative operations of the TA
for calculation of the auxiliary blocks and the solution is given in Tab.~\ref{tab0}.
According to~\cite{Gulin}, we assume that one needs exactly $n^{3}$ multiplicative operations for multiplication
and exactly $n^{3}$ multiplicative operations for inversion.
The inversion is performed by the LU decomposition which takes $n^{3}/3$ multiplicative operations
and then by solving a matrix system with unity matrix in the RHS which, in turn, takes $2n^{3}/3$ multiplicative operations.
The uncertainty of our estimation is $O(n)$, that can be ignored for blocks of the block-tridiagonal matrix of size $n>10$.
As a result, it is straightforward to obtain that the total number of serial multiplicative operations of the TA is~\cite{Isaacson}
\begin{equation*}
\label{thomas}
\text{Mo}_{\text{TA}}=(3N-2) (n^{3}+n^{2}l).
\end{equation*}

\begin{table}[!h]
\begin{center}
\begin{tabular}{cc}
\hline \hline
Calculation of & Number of multiplicative operations \\
\hline \hline
$\mathcal{G}_{i}$, $i=1,\ldots,N$ & $(2N-1)n^3$ \\
$\mathcal{A}_{i}$, $i=1,\ldots,N-1$ & $(N-1)n^3$ \\
$\mathcal{B}_{i}$, $i=1,\ldots,N$ & $(2N-1)n^2 l$ \\
$X_{i}$, $i=N,\ldots,1$ & $(N-1)n^2 l$ \\
\hline \hline
Total & $(3N-2) (n^{3}+n^{2}l)$ \\
\hline \hline
\end{tabular}
\caption{\label{tab0} Number of multiplicative operations for the each part of the TA as well as the total amount. The computational costs of the elementary operations are estimated according to Refs.~\cite{Isaacson,Gulin,Linpack}.}
\end{center}
\end{table}

The estimation of the number of multiplicative operations of the DM is more difficult.
In our analysis, we will firstly consider the DM without recursivity.
It means that the solution of the subsysems in Eq.~(\ref{e231}) and supplementary matrix equation~(\ref{e232}) is performed by the TA.
Later, we will study recursive DM calls for these parts of the method and consider consequences of these improvements.

\subsection{Decomposition method without recursivity}

\begin{table}[!h]
\begin{center}
\begin{tabular}{cc}
\hline \hline
Calculation of & Number of multiplicative operations \\
 & for each $k=1,\ldots,M$ \\
\hline \hline
 & $k=1$:~$(4N_{k}-2)n^{3}+(3N_{k}-2)n^{2}l$ \\
 solution of Eq.~(\ref{e231}) & $1 < k < M$:~$(7N_{k}-4)n^{3}+(3N_{k}-2)n^{2}l$ \\
 & $k=M$:~$(6N_{k}-4)n^{3}+(3N_{k}-2)n^{2}l$ \\
\hline
 multiplications $W^{k}_{L} Z^{k}$ & $k=\{1,M\}$:~$n^{3}+n^{2}l$ \\
 and $W^{k}_{L} z^{k}$ in Eq.~(\ref{e2311}) & $1 < k < M$:~$4n^{3}+2n^{2}l$ \\
\hline
 solution of Eq.~(\ref{e232}) & $(3M-5)(n^{3}+n^{2}l)$ (independently on $k$) \\
\hline
 multiplications in Eq.~(\ref{endSolution}) & $k=\{1,M\}$:~$N_{k} n^{2}l$ \\
  & $1 < k < M$:~$2N_{k} n^{2}l$ \\
\hline \hline
 & $\Big( (4N_{1}-1)n^{3}+(4N_{1}-1)n^{2}l \Big) + $ \\
Total & $\Big( (6N_{M}-3)n^{3}+(4N_{M}-1)n^{2}l \Big) +$ \\
 & $+ \sum_{k=2}^{M-1} \Big( 7N_{k}n^{3} + 5N_{k}n^{2}l \Big) + (3M-5) (n^{3}+n^{2}l)$ \\
\hline \hline
\end{tabular}
\caption{\label{tab2} Number of multiplicative operations for the each stage of the DM as well as the total amount. The computational costs of the elementary operations are estimated according to Refs.~\cite{Isaacson,Gulin,Linpack}.}
\end{center}
\end{table}

The number of multiplicative operations for each stage of the DM is summarized in Tab.~\ref{tab2}.
For solving one independent subsystem with different vectors in RHS using the TA, at the forward stage one calculates the auxiliary blocks $\mathcal{A}_{i}$ for $i=1,\ldots, N_{k}-1$ and $\mathcal{G}_{i}$ for $i=1,\ldots, N_{k}$.
Here $N_{k}$ is the number of blocks on the diagonal of the $k$-th subsystem.
It takes $(3N_{k}-2)n^{3}$ multiplicative operations.
These calculations are general for both equations in~(\ref{e231}).
The auxiliary blocks $\mathcal{B}_{i}$ are calculated for each RHS separately.
Since the second equation in~(\ref{e231}) has the complete RHS, it takes $(2N_{k}-1)n^{2}l$ operations to compute all $\mathcal{B}_{i}$.
The sparse structure of the RHS of the first equation in~(\ref{e231}) leads to
$(2N_{k}-1)n^{3}$ or $n^{3}$ multiplicative operations for the case
when the vector in RHS has only the top nonzero block or only bottom nonzero block, respectively.
At the backward stage, the number of multiplicative operations is $(N_{k}-1)n^{3}$ and $(N_{k}-1)n^{2}l$
for the first and second equation in~(\ref{e231}).
In total, the first subsystem ($k=1$) takes $(4N_{k}-2)n^{3}+(3N_{k}-2)n^{2}l$ multiplicative operations,
the last one ($k=M$) takes $(6N_{k}-4)n^{3}+(3N_{k}-2)n^{2}l$,
the intermediate ones $k=2,\ldots,M-1$ take $(7N_{k}-4)n^{3}+(3N_{k}-2)n^{2}l$.

The calculation of the products $W^{k}_{L}Z^{k}$ and $W^{k}_{L}z^{k}$ takes $4n^{3}$ and $2n^{2}l$ multiplicative operations for $1<k<M$,
whereas for $k=\{1,M\}$ the products take only $n^{3}$ and $n^{2}l$ multiplicative operations.
The construction of matrices~(\ref{e2311}) includes only additive operations and, therefore,
their contribution is ignored.

The solution of the supplementary matrix system~(\ref{e232}) of the block size $M-1$ takes $(3M-5)(n^{3}+n^{2}l)$ multiplicative operations.

The last stage, namely obtaining the remaining unknown vector~(\ref{endSolution}) from the solution of the supplementary matrix system,
takes $2N_{k}n^{2}l$ and $N_{k}n^{2}l$ multiplicative operations for $1<k<M$ and $k=\{1,M\}$, respectively.


As a result, the total number of multiplicative operations of the DM is given as
\begin{equation}
\begin{array}{c}
\displaystyle \Big( (4N_{1}-1)n^{3}+(4N_{1}-1)n^{2}l \Big) + \Big( (6N_{M}-3)n^{3}+(4N_{M}-1)n^{2}l \Big) + \\
\displaystyle + \sum_{k=2}^{M-1} \Big( 7N_{k}n^{3} + 5N_{k}n^{2}l \Big) + (3M-5) (n^{3}+n^{2}l),
\end{array} 
\label{DMtotal}
\end{equation}
where
the last term corresponds to operations needed to solve supplementary matrix equation~(\ref{e232})
and $N_{k}$ is the size of $k$-th subsystem.
It is worth noting that $N_{k}$ can be different for different $k$ and satisfies
the only restriction
$$
\sum_{k=1}^{M} N_{k}=N-(M-1).
$$

\subsection{Recursive call of the decomposition method}

The recursive call of the DM implies using the DM also for solving the supplementary matrix equation~(\ref{e232}) and
the independent subsystems~(\ref{e231}).
We will firstly consider calling the DM for solving the supplementary matrix equation and secondly
for solution of the independent subsystems.
After that, the formulas for the case of calling the DM both for solution of the independent subsystems and the supplementary matrix equation
will be derived.

\subsubsection{Solution of the supplementary matrix equation}

The recursive application of the DM for solving the supplementary matrix equation of size $M-1$, constructed on matrix $\mathbf{H}$~(\ref{e232}),
leads to the change of the number of multiplicative operations in the general scheme, see Tab.~\ref{tab2}.
In particular, if we divide the supplementary matrix equation into $m$ subsystems, then the last term in Eq.~(\ref{DMtotal}) becomes
\begin{equation}
\begin{array}{c}
\displaystyle \Big( (4M_{1}-1)n^{3}+(4M_{1}-1)n^{2}l \Big) + \Big( (6M_{m}-3)n^{3}+(4M_{m}-1)n^{2}l \Big) + \\
\displaystyle + \sum_{j=2}^{m-1} \Big( 7M_{j}n^{3} + 5M_{j}n^{2}l \Big) + (3m-5) (n^{3}+n^{2}l),
\end{array} 
\label{DMrec1}
\end{equation}
where $M_{j}$ is the size of $j$-th subsystem of the matrix equation~(\ref{e232}).
If the subsystems are of equal size, then $M_{j}=(M-m)/m=M_{1}$ for $j=1,\ldots,m$.

\subsubsection{Solution of the independent subsystems}
The solution of the independent subsystems in Eq.~(\ref{e231}) can be done by calling the DM for each such subsystem.
If we divide $k$-th subsystem ($k=1,\ldots,M$) into $J_{k}$ subsubsystems, then
for solving Eq.~(\ref{e231})
for $k=1$
the number of multiplicative operations equals to
\begin{equation*}
\begin{array}{c}
\displaystyle \left( (5L^{1}_{1}-1)n^{3} + (4L^{1}_{1}-1)n^{2}l \right) + \left( (8L_{J_{1}}^{1}-2)n^{3} + (4L^{1}_{J_{1}}-1)n^{2}l \right) + \\
\displaystyle \sum_{i=2}^{J_{1}-1} \left[ L^{1}_{i} \left( 9n^{3}+5n^{2}l \right) \right] + \left( 4J_{1}-6 \right) n^{3} + \left( 3J_{1}-5 \right) n^{2}l
\end{array} 
\label{DMrec121}
\end{equation*}
and for $k=M$
\begin{equation*}
\begin{array}{c}
\displaystyle \left( (8L^{M}_{1}-2)n^{3} + (4L^{M}_{1}-1)n^{2}l \right) + \left( (7L_{J_{M}}^{M}-3)n^{3} + (4L^{M}_{J_{M}}-1)n^{2}l \right) + \\
\displaystyle \sum_{i=2}^{J_{M}-1} \left[ L^{M}_{i} \left( 9n^{3}+5n^{2}l \right) \right] + \left( 6J_{M}-10 \right) n^{3} + \left( 3J_{M}-5 \right) n^{2}l,
\end{array} 
\label{DMrec1212}
\end{equation*}
where $L^{k}_{i}$ is the block-size of the $i$-th subsubsystem for the $k$-th subsystem.
For $1<k<M$ the analogous number is
\begin{equation*}
\begin{array}{c}
\displaystyle \left( (9L^{k}_{1}-2)n^{3} + (4L^{k}_{1}-1)n^{2}l \right) + \left( (9L_{J_{k}}^{k}-2)n^{3} + (4L^{k}_{J_{k}}-1)n^{2}l \right) + \\
\displaystyle \sum_{i=2}^{J_{k}-1} \left[ L^{k}_{i} \left( 11n^{3}+5n^{2}l \right) \right] + \left( 7J_{k}-11 \right) n^{3} + \left( 3J_{k}-5 \right) n^{2}l.
\end{array} 
\label{DMrec122}
\end{equation*}
In order to calculate the total number of multiplicative operations one needs to add multiplications for calculations of $W^{k}_{L} Z^{k}$ and $W^{k}_{L} z^{k}$,
multiplications in Eq.~(\ref{endSolution}), sum up it over $k$, and to add multiplications for solution of Eq.~(\ref{e232}).

\begin{figure}[t!hp]
\centerline{\includegraphics[width=0.5\textwidth, angle=-90.0]{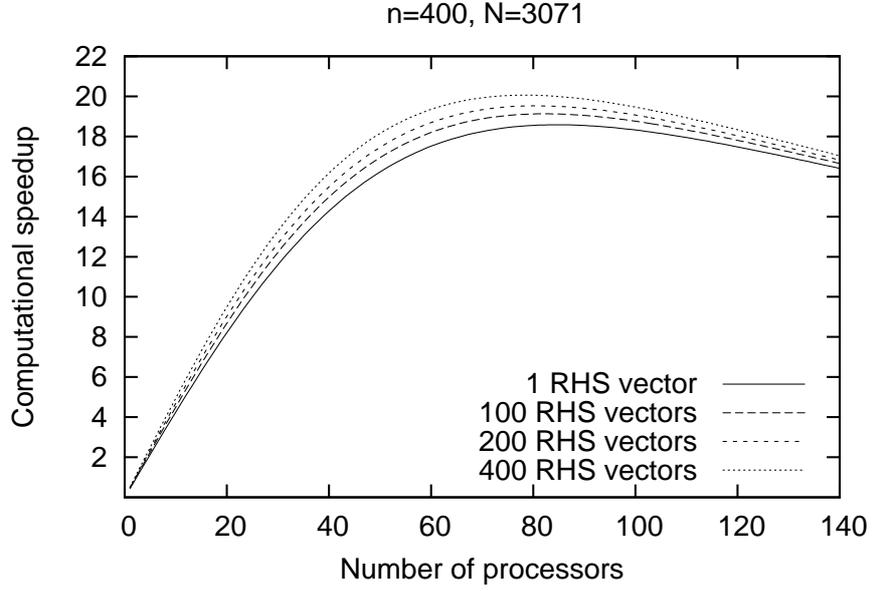}}
\caption{The analytical estimation of the computational speedup of the DM with respect to the TA as a function of the number of parallel processors used for computation. The initial parameters of the matrix are following: the size of each block is $n=400$, number of blocks on the diagonal is $N=3071$. The estimations are shown for various numbers of RHS vectors.}
\label{speedup1}
\end{figure}

\section{Computational speedup}

In order to estimate the computational speedup of the DM with respect to the TA,
one needs to study the relation of overall times of calculations for DM and for the TA.
Since the DM can be executed in parallel on $P$ processors,
in addition to the time of serial calculations, the overall time includes the largest time of computation among all parallel processors.
The overall computation time is directly related to the number of total serial multiplicative operations.
Therefore, to estimate the computational speedup, we evaluate the ratio of the serial multiplicative operations for the TA and for the DM
\begin{equation*}
\label{Speedup}
S=\frac{\text{Mo}_{\text{TA}}}{\text{Mo}_{\text{DM}}}.
\end{equation*}
Below, we will firstly study the computational speedup of the DM without recursivity and after that with it.

\subsection{Decomposition method without recursivity}
The standard usage of the DM includes the TA for solution of the equations for independent subsystem as well as for supplementary problem constructed with coupling matrix $\mathbf{H}$.
Let us consider following cases:
%
\begin{figure}[t!hp]
\centerline{\includegraphics[width=0.5\textwidth, angle=-90.0]{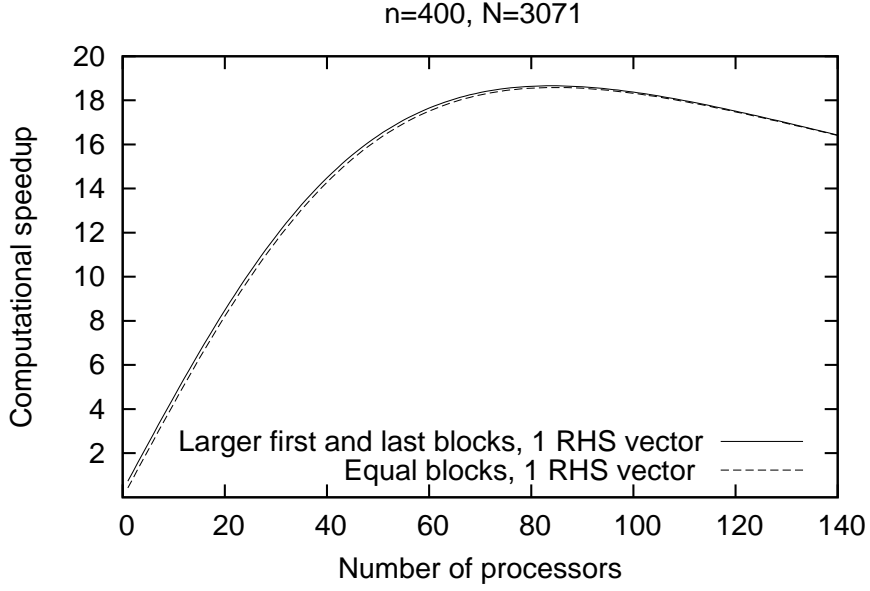}}
\caption{The analytical estimation of the computational speedup of the DM with respect to the TA as a function of the number of parallel processors used for computation. The initial parameters of the matrix are following: the size of each block is $n=400$, number of blocks in the diagonal is $N=3071$. The computational speedup slightly increases if one defines the first and last subsystems are larger than others and, simultaneously, equal computation time for each parallel processor.}
\label{speedup2}
\end{figure}
%
\begin{itemize}
  \item The first one is when the number of subsystems equals to the number of parallel processors $(M=P)$ and all subsystems are of equal size $N_{k}=(N-P+1)/P$ for $k=1,\ldots,M$. In this case
$$
\text{Mo}_{\text{DM}} = \left( \frac{N-P+1}{P} \right) (7n^{3}+5n^{2}l) + (3P-5) (n^{3}+n^{2}l)
$$
and the computational speedup $S$ is given by
\begin{equation}
\label{SpeedupMP}
S=\frac{3N-2}{3P-5+\left(5+ \frac{2}{1+l/n}\right) \left( \frac{N-P+1}{P} \right) } > \frac{1}{\frac{7}{3P}+\frac{P}{N}}.
\end{equation}
So, the computational speedup increases linearly $S\sim 3P/7$ with respect to number of processes $P$ for $P \ll N$.
The computational speedup depends only on the ratio  $l/n$.
This dependence as well as linear growth of the speedup are shown in Fig.~\ref{speedup1} for $n=400$ and $N=3071$.
One can see that the speedup slightly increases if number of vectors $l$ in RHS approaches $n$.
Moreover, it is clearly seen from the plot, that the speedup decreases when the number of processors (and the logical subsystems) becomes large.
The reason for this effect is the growth of the serial part of the DM, i.e. of the supplementary matrix system.
The maximum computational speedup is achieved for the number of processors
\begin{equation*}
\label{MAXSpeedP}
P=\sqrt{ \frac{(N+1)}{3} \left( 5+\frac{2}{1+l/n} \right) }
\end{equation*}
which is close to $\sqrt{7(N+1)/3}$ for $l \ll n$.

 \item The second case is when the number of subsystems equals to the number of processors $(M=P)$ and subsystems are not of equal size $N_{k}$ for $k=1,\ldots,M$.
Since the first and last processors perform less operations than other ones,
it is optimal to define the equal time for each subsystem by varying $N_{k}$ for $k=1,\ldots,M$.
Let us suppose the sizes of the first and last subsystems to be increased in $\alpha$ and $\beta$ times, respectively.
Then, the equal sizes of $k$th subsystems for $k=2,\ldots,M-1$ is
$$
N_{k} = \frac{N-(P-1)}{P-2+\alpha+\beta}.
$$
The computational speedup in this case is defined as
\begin{equation}
\label{SpeedupMPNk}
S=\frac{3N-2}{3P-5+\left(5+ \frac{2}{1+l/n}\right) \left( \frac{N-P+1}{P-2+\alpha+\beta} \right) }.
\end{equation}
One can easily derive from Tab.~\ref{tab2} that $N_{1} \in [(7N_{k}+1)/4,(6N_{k}+1)/4]$
and $N_{M} \in [(7N_{k}+3)/6,(12N_{k}+4)/10]$ for $1 \le l \le n$.
Therefore, for $N_{k} \gg 3$ and $l \ll n$, one approximately obtaines $\alpha \approx 7/4$ and $\beta \approx 7/6$.
Fig.~\ref{speedup2} shows that the computational speedup for $l \ll n$ increases slightly if one defines larger first and last blocks and simultaneously equal computation time for each parallel processor.
For $l \sim n$, the similar almost negligible difference is kept.
 \item The last case is when the number of subsystems is proportional to the number of processors $(M=iP)$ for $i \ge 2$ and subsystems are of the same size as in the previous case.
In this case, each processor is feeded by a queue of $i$ subsystems which are executed sequentially.
Changing
Eq.~(\ref{SpeedupMPNk}) allows one to estimate the speedup in this case:
\begin{equation*}
\label{SpeedupMiPNk}
S=\frac{3N-2}{3iP-5+i\left(5+ \frac{2}{1+l/n}\right) \left( \frac{N-iP+1}{iP-2+\alpha+\beta} \right) }.
\end{equation*}
The additional factor $i$ in front of brackets in the denominator comes from the fact that the queue for each processor consists of $i$ subsystems.
The result for $M=P,2P,3P$ is shown in Fig.~\ref{speedup3}.
It is clearly seen that the maximum speedup is diminished with increase of $i$.
Nevertheless, for small number of processors the linear growth of speedup is kept independently of $i$.
For larger $P$, the serial part of the DM drastically increases, especially for larger $i$.
This fact leads to the considerable decrease of the speedup.
\end{itemize}
The studied cases bring us to the conception for application of the DM.
One should follow the first or second considered case:
subsystems should be of equal size or to have equal execution time.
For both cases the number of subsystems should be equal to number of processors.
The equal execution time is achieved by choosing the size of the first and last subsystems
to be about $7/4$ and $7/6$ of the size of each other subsystem.
Since the difference between described cases is almost negligible, if the size of each subsystem is large,
the initial matrix can be simply divided to equal subsystems.
The computational speedup with respect to the sequential TA is shown in Fig.~\ref{speedup1}.

\begin{figure}[t!hp]
\centerline{\includegraphics[width=0.5\textwidth, angle=-90.0]{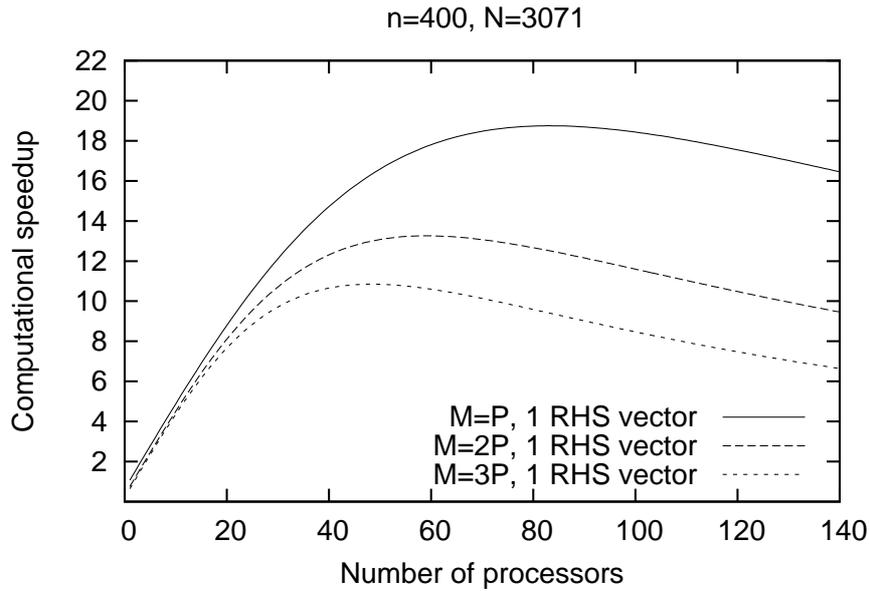}}
\caption{The analytical estimation of the computational speedup of the DM with respect to the TA as a function of the number of parallel processors used for computation. The initial parameters of the matrix are following: the size of each block is $n=400$, number of blocks in the diagonal is $N=3071$. The computational speedup is presented for the case when the first and last subsystems are larger than others and, simultaneously, equal computation time for each parallel processor is defined. Different curves indicate different number of independent subsystems, $M$, which is proportional to number of processors $P$.}
\label{speedup3}
\end{figure}

\subsection{Recursive call of the decomposition method}

We will firstly consider calling the DM for solving only the supplementary matrix equation and
secondly calling it for solution of the independent subsystems and
thirdly both for solution of the independent subsystems and the supplementary matrix equation.

\subsubsection{Parallel solution of the supplementary matrix equation}

\begin{figure}[t!hp]
\centerline{\includegraphics[width=0.5\textwidth, angle=-90.0]{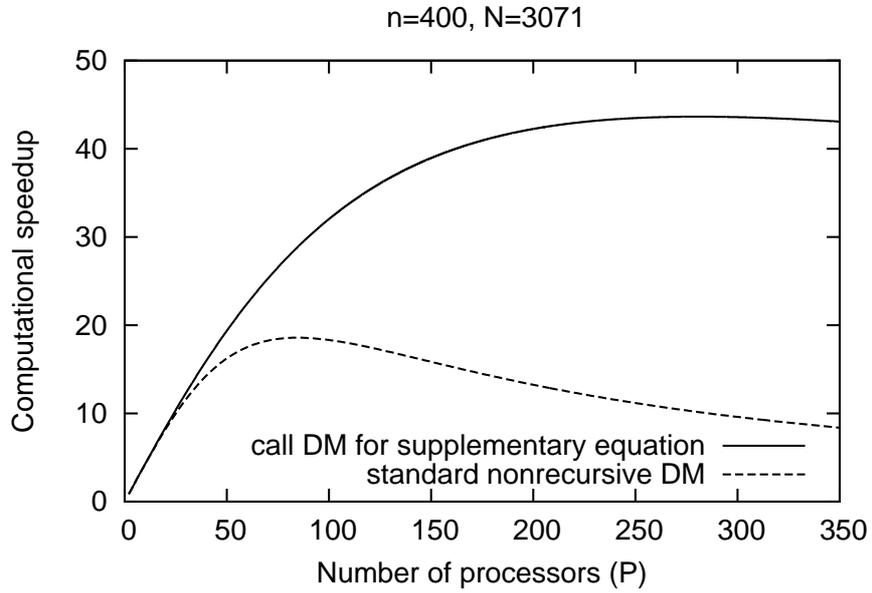}}
\caption{The analytical estimation of the computational speedup of the DM with respect to the TA as a function of the number of parallel processors, $P$, needed for solution of the initial matrix.
The standard application of the DM and the recursive calls for the supplementary matrix system are indicated.
The number of parallel processors, $m$, needed for the supplementary matrix system is given by Eq.~(\ref{SuppMatrM}).}
\label{speeduprec13}
\end{figure}

\begin{figure}[t!hp]
\centerline{\includegraphics[width=0.5\textwidth, angle=0.0]{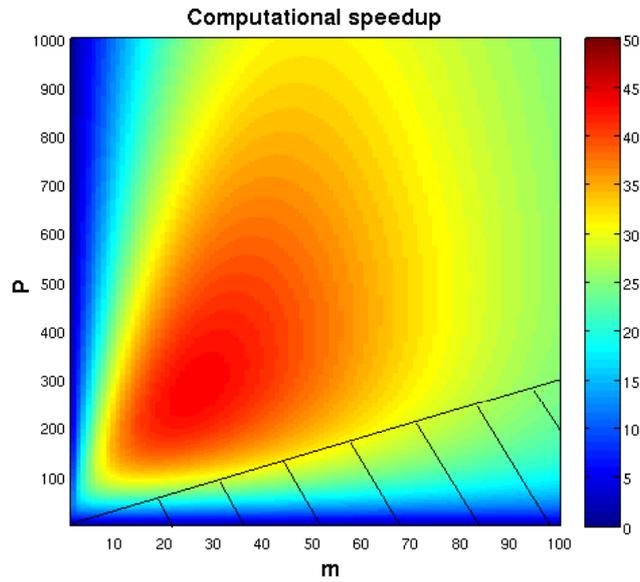}}
\caption{The analytical estimation of the computational speedup of the DM with respect to the TA as a function of the number of parallel processors, $m$, needed for the supplementary matrix system and number of processors, $P$, for solution of the initial matrix by the DM.
The maximum of computational speedup is shown by red color. The bottom dashed area indicates the domain where the condition $m \le P/3$ is not satisfied.}
\label{speeduprec2}
\end{figure}

Based on the previous subsections and particularly Eq.~(\ref{DMrec1}), we now additionally assume that all subsystems $k=1,\ldots,M$ are of equal size $N_{k}=N_{1}$.
As we have seen in previous sections, this assumption does not significantly affect the performance.
If we divide the supplementary matrix equation~(\ref{e232}) into $m$ independent equal subsystems to apply on each of them the DM,
then the number of serial multiplicative operations is given as
\begin{equation}
\text{Mo}_{\text{DM}} = \Big( 7N_{1}n^{3} + 5N_{1}n^{2}l \Big) + \Big( 7M_{1}n^{3} + 5M_{1}n^{2}l \Big) + (3m-5) (n^{3}+n^{2}l), \\
\label{DMrec2}
\end{equation}
where $M_{j}=(M-m)/m=M_{1}$ for $j=1,\ldots,m$ is the size of each subsystem.
As a result, for $M=P$, the following
cases can be considered:
\begin{itemize}
  \item The first case is if $N$ and $P$ are kept fixed. Then, the minimum of Eq.~(\ref{DMrec2}) (and the maximum of the computational speedup) is achieved for
\begin{equation}
\label{SuppMatrM}
m=\sqrt{ \frac{P(7n^{3} + 5n^{2}l)}{3(n^{3} + n^{2}l)} }=\sqrt{ \frac{P}{3} \left( 5+\frac{2}{1+l/n} \right) }.
\end{equation}
Besides, to apply the DM the condition $(P-m)/m \ge 2$ is required. It is satisfied as $P \ge 21$.
If $P<21$, then Eq.~(\ref{SuppMatrM}) does not define the maximum of the speedup, instead it is achieved at the boundary $m=P/3$.
As $l \ll n$, Eq.~(\ref{SuppMatrM}) is reduced to $m \sim \sqrt{{7P}/{3}}$
and the computational speedup $S$ is given as
\begin{equation*}
\label{Speedup521}
S = \frac{3N-2}{7(N+1)/P+2\sqrt{21P}-19}.
\end{equation*}
Fig.~\ref{speeduprec13} shows the considerable increase of the speedup in case of the recursive call of the DM for
the supplementary matrix equation.
  \item The second case is if only $N$ is fixed. This case corresponds to the situation when the number of processors, $P$, is arbitrary
and one has to choose the appropriate number to reach maximum speedup.
One can find parameters of the extremum of Eq.~(\ref{DMrec2}) and show that it is a minimum using the matrix of the second derivatives.
In this case, the minimum of Eq.~(\ref{DMrec2}) is achieved for
\begin{equation}
\label{SuppMatrPM}
P=\sqrt[3]{ \frac{(N+1)^{2}}{3} \left( 5+\frac{2}{1+l/n} \right) },\quad m=\frac{P^{2}}{N+1}.
\end{equation}
Since $m \le P/3$, the parameters~(\ref{SuppMatrPM}) give minimum if $P \le (N+1)/3$, which is usually satisfied.
The computational speedup, calculated for $N=3071$, as a function of $P$ and $m$ is represented in Fig.~\ref{speeduprec2}.
The contour plot shows that the unique maximum (red color) of the speedup exists and is achieved by the given formulas~(\ref{SuppMatrPM}).
\end{itemize}
To summarize the considered cases,
it should be pointed out that the recursive call of the DM for the supplementary matrix equation
leads to remarkable growth of the performance, see Fig.~\ref{speeduprec13}.
Using the described cases, one can choose the parameters of the algorithm based on its own computational facilities and the parameters of the initial matrix in order to achieve the maximum performance, see Fig.~\ref{speeduprec2}.


\subsubsection{Parallel solution of the independent subsystems}
Now we consider parallelization of the solution of each of the $M$ independent subsystems in Eq.~(\ref{e231}).
Solving the supplementary matrix equation~(\ref{e232}) remains to be serial.
We apply the DM for each equal independent subsystem, namely divide each subsystem with $N_{1}$ blocks on the diagonal into
$J$ independent equal subsubsystems.
(One can see that nonequal subsubsystems lead to the considerable reduce of the performance.)
Then, the total number of used processors is $P=MJ$
and the number of serial multiplicative operations is given as
\begin{equation}
\begin{split}
\text{Mo}_{\text{DM}} = \left( 11n^{3} + 5n^{2}l \right) N_{2} + \left( n^{3} (7J-11) + n^{2}l (3J-5) \right)+ \\
+ 4n^{3}+2n^{2}l
+(n^{3}+n^{2}l)(3M-5)
+2n^{2}l N_{1},
\end{split}
\label{DMrec3}
\end{equation}
where $N_{1}=(N-M+1)/M$ and $N_{2}=(N_{1}-J+1)/J$.
In the case when $P$ ($P \le (N+1)/3$) is kept fixed, using the method of Lagrange multipliers~\cite{Bertsekas} one can obtain that the maximum speedup is achieved for the values of parameters
\begin{equation*}
M=\sqrt{\frac{(7n^{3}+3n^{2}l)P+2n^{2}l(N+1)}{3(n^3+n^{2}l)}}, \quad J=\frac{P}{M}.
\label{ParamDMrec3}
\end{equation*}
The resulted computational speedup for $l \ll n$ is shown in
Fig.~\ref{speeduprec5}.
%
It is clearly seen that for small $P$ the speedup behaves as $S \sim 3P/11$.
For larger $P$, the speedup reaches its maximum
and exceeds that obtained when the DM is recursively called for the supplementary matrix equation.

\begin{figure}[t!hp]
\centerline{\includegraphics[width=0.5\textwidth, angle=0.0]{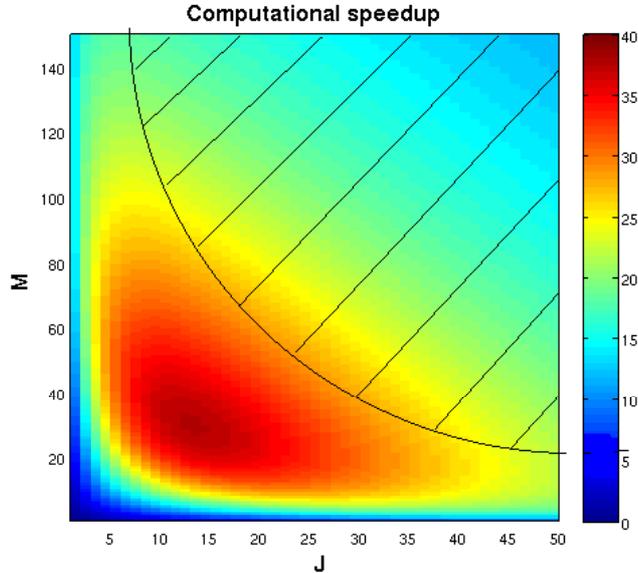}}
\caption{The analytical estimation of the computational speedup of the DM with respect to the TA as a function of the number of subsystems, $M$, the initial matrix is divided into and the number of subsubsystems, $J$, each subsystem is divided into.
The total number of parallel processors involved in computation is $P=MJ$.
The maximum of computational speedup is shown by the dark red color.
The dashed area corresponds to the domain where the condition $P \le (N+1)/3$ is not hold.}
\label{speeduprec4}
\end{figure}

One can also analytically obtain the maximum speedup for the case when $P$ is arbitrary.
It is achieved for the parameter
\begin{equation*}
M=\frac{(N+1)}{J^2} \left( \frac{11n^{3}+5n^{2}l}{7n^{3}+3n^{2}l} \right),
\label{P2DMrec3}
\end{equation*}
and $J$ obtained as a real positive solution of the quartic equation
\begin{equation}
J^{4}+\left( \frac{11n^{3}+5n^{2}l}{2n^{2}l} \right) J^{3}
- (N+1) \frac{3(n^{3}+n^{2}l) \left( 11n^{3}+5n^{2}l \right)^{2}}{\left( 7n^{3}+3n^{2}l \right)^{2} 2n^{2}l} = 0.
\label{P3DMrec3}
\end{equation}
However, in practice, it is much simpler to solve Eq.~(\ref{P3DMrec3}) numerically if the size of the block, $n$, number of RHS vectors, $l$, and $N$ are known.
An example of reached speedup for the case when $n=400$, $l=1$, $N=3071$ as a function of $M$ and $J$ is shown in Fig.~\ref{speeduprec4}.
The dashed area corresponds to the domain where the condition $P \le (N+1)/3$ is not hold and the parallel solution of the independent subsystems is not possible.

\subsubsection{Parallel solution of both the independent subsystems and the supplementary matrix equation}
If we apply the DM to both the independent subsystems and the supplementary matrix equation,
then the number of serial multiplicative operations is the combination of Eq.~(\ref{DMrec2}) and Eq.~(\ref{DMrec3}).
Keeping the notation of the previous case and assuming also that $P=MJ$, the number of serial multiplicative operations is given as 
\begin{equation*}
\begin{split}
\text{Mo}_{\text{DM}} = \left( 11n^{3} + 5n^{2}l \right) N_{2} + \left( n^{3}(7J-11) + n^{2}l (3J-5) \right)+ \\
+ 4n^{3}+2n^{2}l
+(7n^{3}+5n^{2}l) \left( \frac{M-m}{m} \right)
+(n^{3}+n^{2}l)(3m-5)
+2n^{2}l N_{1},
\end{split}
\label{DMrec4}
\end{equation*}
where $N_{1}=(N-M+1)/M$, $N_{2}=(N_{1}-J+1)/J$ and $m$ is the number of independent subsystems
to which the supplementary matrix equation is divided into.

\begin{figure}[t!hp]
\centerline{\includegraphics[width=0.5\textwidth, angle=-90.0]{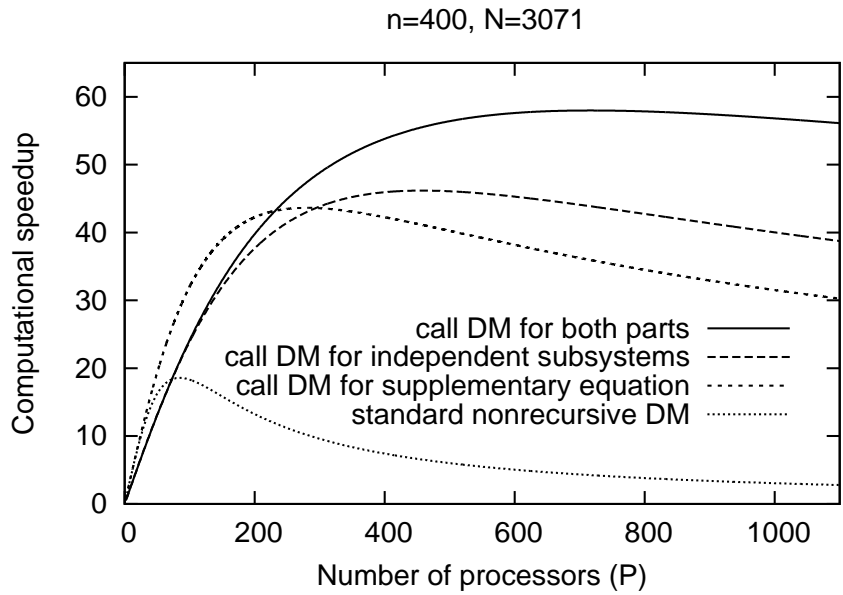}}
\caption{The analytical estimation of the computational speedup of the DM with respect to the TA as a function of the overall number of processors, $P$, needed for parallel computation.
The four cases described in the text are shown.}
\label{speeduprec5}
\end{figure}

Keeping $P$ ($P \le (N+1)/3$) to be fixed, one can obtain the maximum computational speedup if the parameters are
\begin{equation*}
\begin{split}
M = \frac{1}{\sqrt[3]{3 \left( 5+\frac{2}{1+l/n} \right)}} \left( 3P+\frac{4P}{1+l/n}+\frac{2l(N+1)}{n+l} \right)^{2/3}, \\
m = \sqrt{\frac{M (7n^{3}+5n^{2}l)}{3(n^{3}+n^{2}l)}} =
\frac{1}{\sqrt[3]{9}} \left( 3P+\frac{4P}{1+l/n}+\frac{2l(N+1)}{n+l} \right)^{1/3} \left( 5+\frac{2}{1+l/n} \right)^{1/3}
\end{split}
\label{P1DMrec4}
\end{equation*}
and obviously $J=P/M$.
The comparison of the computational speedup for this case with previous cases is shown in Fig.~\ref{speeduprec5}.
It is clearly seen that the computational speedup in this case is larger than in previous cases.
Unfortunately, this large speedup
is achieved only for large total number of involved parallel processors, $P$, that is practically not feasible.

If the total number of involved processors $P$ is free, then the maximum computational speedup is achieved
for values of the parameters
\begin{equation}
\begin{split}
M = \frac{(N+1)}{J^2} \left( \frac{11n^{3}+5n^{2}l}{7n^{3}+3n^{2}l} \right), \quad
m = \sqrt{\frac{M (7n^{3}+5n^{2}l)}{3 (n^{3}+n^{2}l)} }
\end{split}
\label{P2DMrec4}
\end{equation}
and $J$ is the real positive solution of the cubic equation
\begin{equation}
J^{3}+\left( \frac{11n^{3}+5n^{2}l}{2n^{2}l} \right) J^{2}
- \frac{\sqrt{N+1}}{(2n^{2}l)} \left(\frac{11n^{3}+5n^{2}l}{7n^{3}+3n^{2}l}\right)^{3/2} \sqrt{3(7n^{3}+5n^{2}l)(n^{3}+n^{2}l)} = 0.
\label{P3DMrec4}
\end{equation}
In Fig.~\ref{speeduprec5}, the parameters of the analytical maximum provided by formulas~(\ref{P2DMrec4},\ref{P3DMrec4}) are $M=106$, $J=7$, $m=16$.

\section{Validation of the analytical results}
\begin{figure}[t!hp]
\centerline{\includegraphics[width=0.5\textwidth, angle=-90.0]{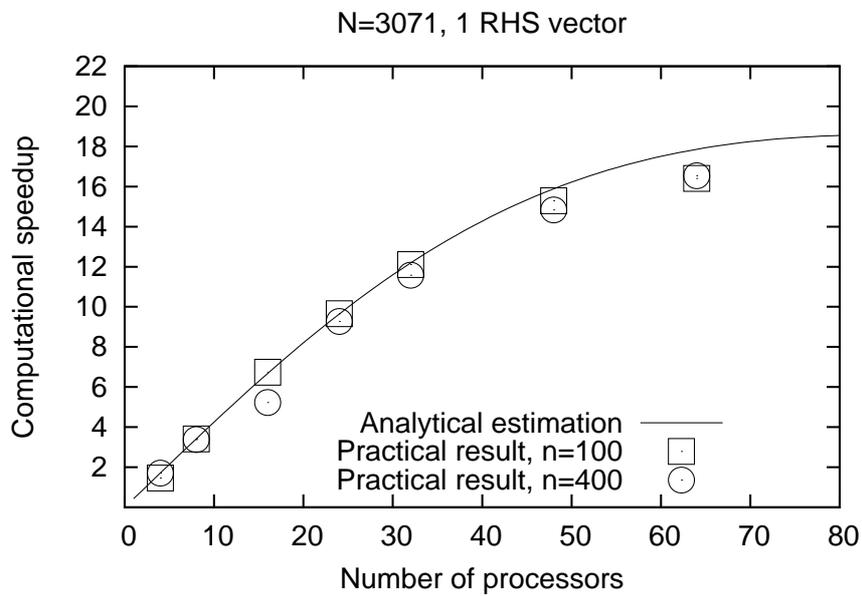}}
\caption{The analytical estimation (solid line) of the computational speedup of the DM with respect to the TA as a function of the number of parallel processors used for computation as well as the speedup obtained in the validation studies (empty squares and circles). The parameters of the matrix are following: the size of each block is $n=100,400$, number of blocks in the diagonal is $N=3071$, only one RHS vector is used.}
\label{speedupvalidation1}
\end{figure}

Described analytical estimations have been validated using the SMP system with
64 processors (Intel Xeon CPU X7560 2.27GHz) and 2TB of shared operative memory.
The Red Hat Enterprise Linux Server 6.2 and GCC 4.4.6 (20110731) compiler are installed in the system.
The DM has been implemented as an independent program written in C with
calls of the corresponding LAPACK 3.5.0 Fortran subroutines~\cite{Lapack350}.
The parallelization has been done using the OpenMP 3.0~\cite{OpenMP}.

In validation studies of the computational speedup~(\ref{SpeedupMP}) only the solution time has been taken into account.
This time includes time needed for computation itself as well as time needed for possible memory management during the solution.
The time for generation of the initial blocks is ignored.
The initial blocks have been generated by the well known discretization
of the two dimensional Laplace operator in the Faddeev equations (see Section 6)
with filling of zero elements by relatively small random values.
This procedure guarantees that the generated matrices are well-posed.

We experimentally estimated the computational speedup for the nonrecursive case when each subsystem~(\ref{e231}) has an equal size.
In Fig.~\ref{speedupvalidation1}, we took the size of a block to be $n=100,\, 400$ and
we show the maximum speedup obtained in a series of experiments with fixed number of processors.
It should be noted that the achieved speedup was equal or less than the analytical one.
The averaged value of the experimentally obtained speedups is smaller than the shown values by $3\%-5\%$.
This difference may be attributed to the nonideality of the memory management of the computational system,
additional system processes running simultaneously and affecting our task,
and hyper-threading of the processors.
These issues are clearly seen for the case when the number of parallel processors is $P=48$ and especially $P=64$.
For these cases, the maximum achieved speedup is considerably less than the analytical one.
%

Nevertheless, the general trend of the analytical result
for the equal subsystems
is clearly confirmed in the validation studies.
The practical behavior of the computational speedup for other cases can also be observed in the computational systems with more processors.
%
%

\section{Application}

We apply the DM to the numerical solution of the boundary value problem (BVP) arised from the three-body scattering problem.
One of the rigorous approaches for treating
the three-body scattering problems above the breakup threshold
is based on
the configuration space Faddeev formalism~\cite{FM}.
It reduces the scattering problem
to the BVP by implementing appropriate boundary conditions.
The
boundary conditions have been introduced by S.~P.~Merkuriev~\cite{M}
and their new representation has recently been constructed in Ref.~\cite{BY}.
After discretization of the BVP at some grid we come to the matrix equation of interest
and then apply the DM.
Below, we will describe key points of the neutron-deuteron ($nd$) scattering problem,
computational scheme, and provide some results.

\subsection{Statement of the problem}
The neutron-deuteron system under consideration is described by the differential Faddeev equations~\cite{FM}.
The $s$-wave equations for the radial part of the Faddeev wave function component appear
after projection onto the states with zero orbital momentum in all pairs of the three-body system.
These $s$-wave integro-differential Faddeev equations for the intrinsic angular momentum $3/2$ in the Cartesian coordinates are given by one equation~\cite{Kuperin,Filikhin}
\begin{equation}
\label{e020}
\left(-\frac{\partial^{2}}{\partial x^{2}}-\frac{\partial^{2}}{\partial y^{2}}+V(x)-E\right)U(x,y)=\frac{1}{2}V(x)
\int\limits_{-1}^{1} d\mu \frac{xy}{x'y'} U(x',y'),
\end{equation}
where
\begin{equation*}
\label{e0710}
x'=\left(\frac{x^{2}}{4}-\frac{\sqrt{3}}{2}xy\mu+\frac{3y^{2}}{4}\right)^{1/2}, \quad
y'=\left(\frac{3x^{2}}{4}+\frac{\sqrt{3}}{2}xy\mu+\frac{y^{2}}{4}\right)^{1/2},
\end{equation*}
and $\mu=\cos{(\hat{x},\hat{y})}$.
The solution of the $s$-wave Faddeev equation~(\ref{e020}) for the energy of the system above the breakup threshold ($E>0$) and for the short-range two-body potential $V(x)$ should
satisfy
the boundary condition~\cite{M}
\begin{equation}
\label{e040c}
U(x,y)\sim\varphi(x)\left(\sin{qy}+a_{0}(q)\exp{iqy}\right)+A(\theta,E)\frac{\exp{i\sqrt{E}\rho}}{\sqrt{\rho}},
\end{equation}
where $\rho=\sqrt{x^{2}+y^{2}}$, $\tan\theta={y/x}$, as $\rho \to \infty$.
The conditions $U(x,0)=U(0,y)=0$ guarantee the regularity of the solution at zero.
The energy and the relative momentum of the neutron, $q$, are associated with the energy of the two-body ground state, defined by the Schr\"{o}dinger equation
\begin{equation}
\label{e0900}
\left(-\frac{d^{2}}{d x^{2}}+V(x)\right)\varphi(x)=\varepsilon\varphi(x),
\end{equation}
by the formula $q^{2}=E-\varepsilon$.
The functions $a_{0}(q)$ and $A(\theta,E)$ to be determined are the binary amplitude and the Faddeev component of the breakup amplitude, respectively.
The integral representations for these functions are of the form \cite{Payne1992}
\begin{equation}
\label{e041a}
a_{0}(q) = \frac{1}{q}\int\limits_{0}^{\infty} \, dy \, \sin{qy} \int\limits_{0}^{\infty} \, dx \, \varphi(x) \, \mathcal K(x,y)
\end{equation}
and
\begin{equation}
\label{e041b}
A(\theta,E) = \sqrt{\frac{2}{\pi\sqrt{E}}} e^{i\pi/4} \int\limits_{0}^{\infty} \, dy \, \sin{(\sqrt{E}\sin{\theta} \; y)} \int\limits_{0}^{\infty} \, dx \, \phi(\sqrt{E}\cos{\theta},x) \, \mathcal K(x,y),
\end{equation}
where $\phi(k,x)$ is the scattering two-body wave function
\begin{equation*}
\label{e042}
\phi(k,x)\underset{x\to\infty}{\to} e^{i\delta(k)} \sin \left(kx+\delta(k) \right)
\end{equation*}
and
\begin{equation*}
\label{e043}
\mathcal K(x,y) = \frac{1}{2} V(x)
\int\limits_{-1}^{1} d\mu \frac{xy}{x'y'} U(x',y').
\end{equation*}

\subsection{Numerical method}
The solution of the BVP with equation~(\ref{e020})
for the ground and excited states as well as for scattering
have been performed by various authors~\cite{Roudnev,RL3}.
The solution method is based on expanding the solution in a basis of the Hermite splines $H$ over $x$ and $y$
\begin{equation*}
\label{HermiteSplineExpansion}
U(x,y)= \sum_{i=0}^{N_{x}} H^{x}_{i}(x) \sum_{j=0}^{N_{y}} H^{y}_{j}(y) \, c_{ij}
\end{equation*}
and allows one to calculate binding energies and Faddeev components.
For scattering, the amplitudes are then reconstructed from the Faddeev components using the integral representations~(\ref{e041a},~\ref{e041b}).
The matrix system is obtained after the discretization of the equation~(\ref{e020})
at the two-dimensional grid.
In addition to the difficult treatment of the boundary conditions~(\ref{e040c}) in case of scattering,
this approach has a computational disadvantage 
consisting in the
relatively irregular and nonband
structure of the matrix of the resulted system of linear equations.
The block-tridiagonal structure does not arise.

Another approach is to rewrite the equations in hyperspherical coordinates~\cite{Simonov,Badalyan,Vin}.
This approach was firstly introduced in Ref.~\cite{M} and has been used by a number of authors~\cite{M1986,Chen,Motovilov,Payne}.
It allows one to obtain the system of linear equations with a block-tridiagonal matrix.
It is more appropriate for establishing the boundary conditions~(\ref{e040c}) as well.
Taking into account the change of the unknown function as
$\mathcal U(\rho,\theta)\equiv \sqrt{\rho}U(x,y)$,
the transformation to the hyperspherical (polar) coordinates $\{\rho,\theta \}$ leads to the following equation:
\begin{equation}
\label{e050a}
\left(-\frac{\partial^{2}}{\partial\rho^{2}}-\frac{1}{4\rho^{2}}-\frac{1}{\rho^{2}} \frac{\partial^{2}}{\partial\theta^{2}}+V
(\rho \cos{\theta})-E\right) \mathcal U(\rho,\theta)
=\frac{2}{\sqrt{3}}V(\rho \cos{\theta})\int\limits_{\theta_{-}(\theta)}^{\theta_{+}(\theta)}
\mathcal U(\rho,\theta') \, d\theta'.
\end{equation}
The integration limits are defined, in turn, as $\theta_{-}(\theta)=|\pi/3-\theta|$ and $\theta_{+}(\theta)=\pi/2-|\pi/6-\theta|$.
%
The boundary condition~(\ref{e040c})
can be represented as~\cite{BY}
\begin{equation}
\label{e070a}
\mathcal U(\rho,\theta)\sim \phi_{0}(\rho|\theta)\left( \mathcal Y_{0}(q\rho) +a_{0}(q) \mathcal H_{0}(q\rho) \right)+
 \sum_{k=1}^{N_{\phi}} \phi_{k}(\rho|\theta) a_{k}(E) \mathcal H_{k}(\sqrt{E}\rho),
\end{equation}
where $\mathcal Y_{0}(t)$ and $\mathcal H_{k}(t)$ are expressed by
\begin{equation*}
\label{e2211}
\mathcal Y_{0}(t)=\sqrt{\frac{\pi t}{2}} \frac{Y_{0}(t)+J_{0}(t)}{\sqrt{2}}, \quad
\mathcal H_{k}(t)=\sqrt{\frac{\pi t}{2}}H_{2k}^{(1)}(t)\exp{i(\pi/4+\pi k)}
\end{equation*}
through the Bessel functions $J_{0},Y_{0}$ and the Hankel functions of the first kind  $H^{(1)}_{k}$~\cite{AS}.
The unknown Faddeev component of the breakup amplitude is expressed in this case as the expansion
\begin{equation*}
\label{e080}
A(\theta,E)=\lim_{\rho\rightarrow\infty} A(\theta,E,\rho)=\lim_{\rho\rightarrow\infty} \sum_{k=1}^{N_{\phi}} a_{k}(E)
\phi_{k}(\rho|\theta),
\end{equation*}
whereas the binary amplitude is given by $a_{0}(q)$.
Here, the functions $\phi_{k}(\rho|\theta)$ form the basis
generated by the eigenvalue problem for operator of the two-body subsystem.
The procedure for extraction of the coefficients $a_{k}$ of the presented expansion is given in Ref.~\cite{BY2014}.

\begin{figure}
\centerline{\includegraphics[width=0.5\textwidth, angle=-90.0]{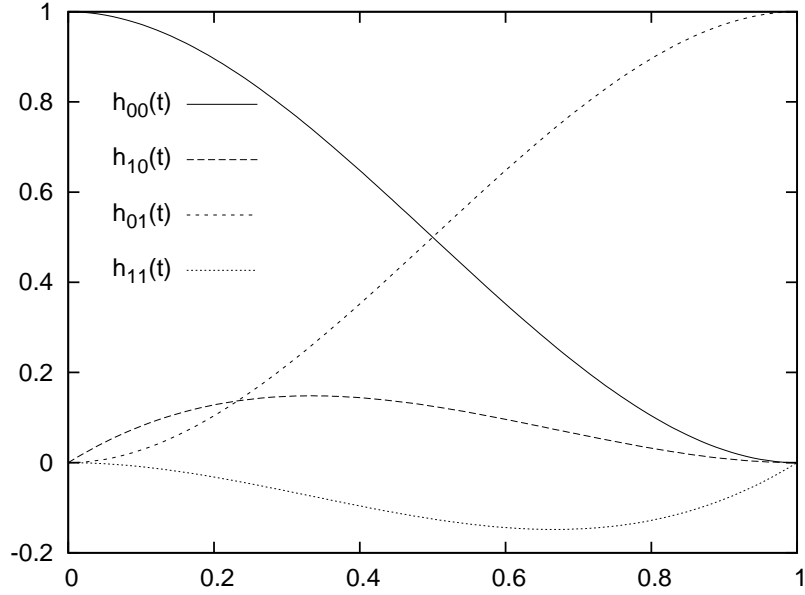}}
\caption{The cubic Hermite splines (\ref{e175}) at the unit interval.}
\label{fig10}
\end{figure}

The two-dimensional BVP defined by
Eq.~(\ref{e050a})
and boundary condition~(\ref{e070a})
 is solved in the hyperspherical coordinates $\{\rho,\theta\}$
due to proper description of the boundary conditions
and appropriate representation of the operator of two-body subsystem in Eq.~(\ref{e050a}).
Therefore, the computational scheme meets the requirements of a good representation of the $\theta$-dependent operator.
As a result, at $N_{\theta}$ intervals over the $\theta$-coordinate the unknown solution is expanded in the basis of cubic Hermite splines $H(\theta)$~\cite{Payne}
\begin{equation*}
\label{e270}
U(\theta,\rho)=\sum_{i=0}^{N_{\theta}}\enskip H^{\rho}_{i}(\theta) c_{i}(\rho),
\end{equation*}
whereas over the $\rho$-coordinate the finite-difference scheme has been taken.
At the unit interval $t\in[0,1]$, the cubic Hermite splines are defined by four formulas
\begin{equation}
\label{e175}
\begin{array}{lcl}
h_{00}(t) &=&2t^{3}-3t^{2}+1,\\
h_{10}(t) &=&t^{3}-2t^{2}+t,\\
h_{01}(t) &=&-2t^{3}+3t^{2},\\
h_{11}(t) &=&t^{3}-t^{2}.
\end{array}
\end{equation}
The splines are shown in Fig.~\ref{fig10}.
They transferred by the linear transformations to the two consecutive intervals of $\theta$-grid.
Parameterizing these intervals by $t\in[-1,1]$, the splines can be written in the following way:
\begin{eqnarray*}
\label{e20200}
H^{1}(t)&=&\left\{ \begin{array}{r@{}ll} -2&t^{3}-3t^{2}+1, & t\in[-1,0) \\
					2&t^{3}-3t^{2}+1, & t\in[0,1] \\
		 \end{array}\right.,\\
H^{2}(t)&=&\left\{ \begin{array}{r@{}ll}  &t^{3}+2t^{2}+t, & t\in[-1,0) \\
					&t^{3}-2t^{2}+t, & t\in[0,1] \\
		 \end{array}\right..
\end{eqnarray*}

In order to obtain the appropriate $\theta$-grid, the specially chosen nonequidistant $x$-grid for operator
of the two-body subsystem
has been used and transformed by the relation
\begin{equation*}
\label{e180}
\theta_{i}(\rho) = \arccos{\frac{x_{i}}{X(\rho)}}, \quad \theta_{i}\in[0,\pi/2].
\end{equation*}
Here, the parameter $X(\rho)$ defines the $x$-coordinate of the right zero boundary condition for some $\rho$.
The $x$-grid consists of two parts: the fixed one at small $x$ and the stretchable one at larger $x$ up to $X(\rho)$.
As $\rho$ increases, $X(\rho)$ grows and the obtained $\theta$-grid becomes more dense near $\pi/2$.
The quality of the $x$-grid and consequently of the $\theta$-grid has been estimated by
a precision of the ground state eigenvalue of the two-body Hamiltonian~(\ref{e0900}).
The spline-expansion of the solution requires using as many as twice of numbers of coefficients in the expansion.
The orthogonal collocation method~\cite{Prenter} with two Gauss knots within one interval is used for discretization of the studied equation.
Over the equidistant $\rho$-grid with the mesh parameter $\Delta\rho=\rho_{m}-\rho_{m-1}$
the second partial derivative of the equation (\ref{e050a}) is approximated
by the second order finite-difference formula
\begin{equation}
\label{e20100}
\frac{\partial}{\partial \rho^{2}} \mathcal U(\rho,\theta) \rightarrow
\frac{\mathcal U(\rho_{m-1},\theta)-2 \, \mathcal U(\rho_{m},\theta)+\mathcal U(\rho_{m+1},\theta)}{(\Delta\rho)^{2}}.
\end{equation}
This approximation generates the block-tridiagonal structure for the matrix of the linear system,
which can be written at the grid
$\{\theta^{(m)}_{j}\}_{j=1}^{2N_{\theta}}$
as follows:
\begin{multline}
\label{e290}
\sum_{i=0}^{N_{\theta}} \Bigg[-\frac{
				H_{i}^{\rho_{m-1}}(\theta_{j}^{(m)}) \, c_i(\rho_{m-1})
				-2H_{i}^{\rho_{m}}(\theta_{j}^{(m)}) \, c_i(\rho_{m})
				+H_{i}^{\rho_{m+1}}(\theta_{j}^{(m)}) \, c_i(\rho_{m+1})
		}{(\Delta\rho)^2}
		+ \Bigg\{-\frac{1}{\rho^{2}_{m}} \frac{\partial^{2}H^{\rho_{m}}_{i}}{\partial^{2}\theta}\left(\theta_{j}^{(m)}\right) + \\
		+ \left(
			V(\rho_{m} \cos{\theta_{j}^{(m)}})-\frac{1}{4\rho_{m}^{2}}-E
		\right)
		 H_{i}^{\rho_{m}}(\theta_{j}^{(m)})
 - \frac{2}{\sqrt{3}} V(\rho_{m}\cos{\theta^{(m)}_{j}})  \int\limits_{\theta_{-}(\theta^{(m)}_{j})}^{\theta_{+}(\theta^{(m)}_{j})} H_{i}^{\rho_{m}}(\theta') \, d\theta' \Bigg\} \, c_i(\rho_{m})
		 \Bigg]
=0
\end{multline}
Here, the index $(m)$ denotes the number of arc $\rho_{m}$ on which the grid $\{\theta^{(m)}_{j}\}_{j=1}^{2N_{\theta}}$ is constructed.
As one can see from Eq.~(\ref{e290}), the resulted system of linear equations has a block-tridiagonal matrix.
Moreover, since the inital equation is integro-differential, the diagonal blocks of the matrix are dense,
whereas the offdiagonal ones are banded.
The RHS supervector of the system is zero
unless the last block-row consisting of terms of the boundary condition~(\ref{e070a}) at the last arc of the hyper-radius.

The proposed numerical method guarantees the precision of the obtained solution of order of $(\Delta\rho)^{2}$
at equidistant grid over $\rho$ and of order of $(\Delta\theta)^{4}$ if one designes the equidistant grid over $\theta$.
The precision over coordinate $\rho$ can be improved up to $(\Delta\rho)^{4}$ if one applies the Numerov method~\cite{Suslov} for the given problem,
that holds the block-tridiagonal structure of the matrix.
The increase of the accuracy over coordinate $\theta$ can be achieved by employing the quintic Hermite splines
\begin{equation*}
\label{eQuinticHermite}
\begin{array}{lcl}
h_{50}(t) &=& -6t^{5}+15t^{4}-10t^{3}+1,\\
h_{51}(t) &=& -3t^{5}+8t^{4}-6t^{3}+t,\\
h_{52}(t) &=& -0.5t^{5}+1.5t^{4}-1.5t^{3}+0.5t^{2},\\
h_{53}(t) &=& 0.5t^{5}-t^{4}+0.5t^{3},\\
h_{54}(t) &=& -3t^{5}+7t^{4}-4t^{3},\\
h_{55}(t) &=& 6t^{5}-15t^{4}+10t^{3}.
\end{array}
\end{equation*}
The downside of the quintic Hermite splines is that the number of collocation point in this case should be increased in more than one and half times comparing to the case of qubic ones.

\subsection{Results of calculations}
\label{sec4}
The calculations have been carried out for various laboratory frame energies, $4 \le E_{lab} \le 42$~MeV,
using the MT~\Rmnum{1}-\Rmnum{3} potential~\cite{Friar1995} for description of the two-body subsystem~(\ref{e0900}).
For this potential, the achieved value of two-body ground state energy is $E_{2b}=-2.23069$~MeV.
The nonequidistant $\theta$-grid with about 500 intervals used for the precise calculations.
Since the orthogonal collocation method~\cite{Prenter} with two Gauss knots within one interval is used, the common size for a matrix of the two-body operator~(\ref{e0900}) is about 1000.
A grid mesh for
the uniform $\rho$-grid was varied from $0.05$~fm to $0.01$~fm.

\begin{figure}[t]%
\begin{tabular}{c}
\begin{minipage}{0.50\linewidth}
\includegraphics[width=1.0\textwidth, angle=0.0]{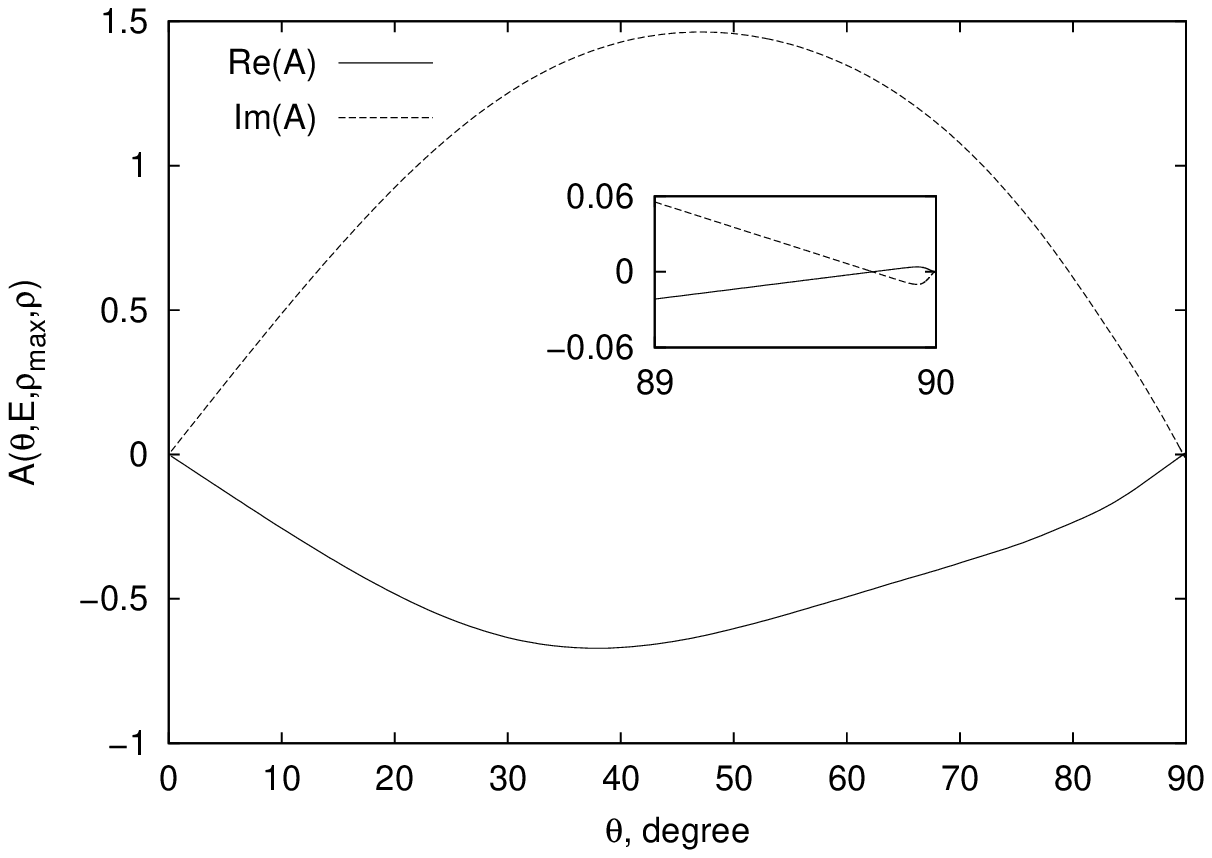}
\end{minipage}
\begin{minipage}{0.50\linewidth}
\includegraphics[width=1.0\textwidth, angle=0.0]{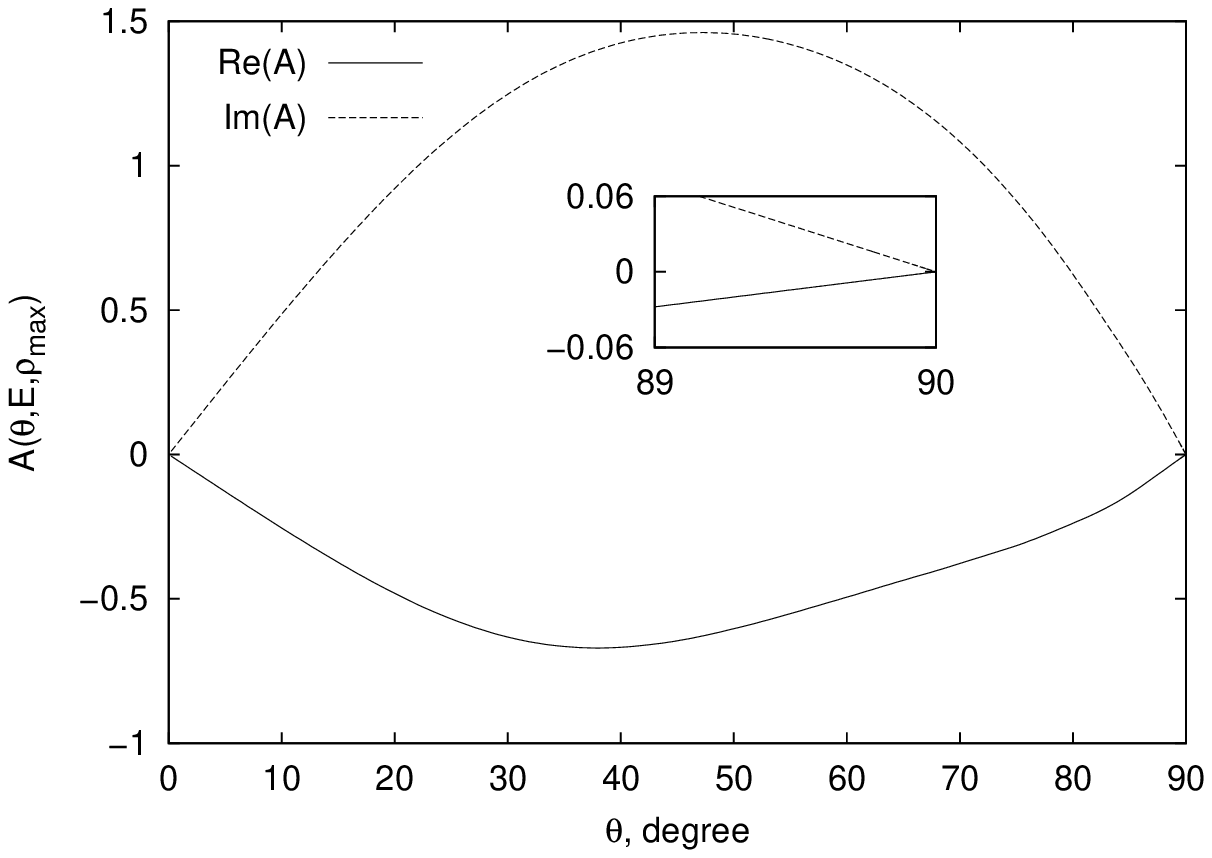}
\end{minipage}
\end{tabular}
\caption{Left panel: the prelimiting breakup amplitude $A(\theta,E,\rho_{max},\rho)$ for $E_{lab}=14.1$~MeV and $\rho=\rho_{max}=1400$~fm.
Right panel: the breakup amplitude $A(\theta,E,\rho_{max})$ for $E_{lab}=14.1$~MeV and $\rho_{max}=1400$~fm.}
\label{fig3040}
\end{figure}

Within the asymptotic approach, the BVPs consisting of the equation~(\ref{e050a})
and the boundary condition~(\ref{e070a})
taken at the hyper-radius $\rho=\rho_{max}$ have been solved.
The expansion coefficients $a_{k}(E,\rho_{max})$ as functions of $\rho_{max}$ have been calculated
and used for reconstructing of the Faddeev component of the breakup amplitude
\begin{equation*}
\label{e240}
A(\theta,E,\rho_{max})=\lim_{\rho\rightarrow\infty} A(\theta,E,\rho_{max},\rho)=
\sum_{k=1}^{N_{\phi}} a_{k}(E,\rho_{max}) \lim_{\rho\rightarrow\infty} \phi_{k}(\rho|\theta).
\end{equation*}
The prelimiting breakup amplitude, $A(\theta,E,\rho_{max},\rho)$,
for $E_{lab}=14.1$~MeV at some finite value of $\rho_{max}$ is shown in Fig.~\ref{fig3040} (left panel).
The breakup amplitude $A(\theta,E,\rho_{max})$ as $\rho\to\infty$ for the same energy is presented in Fig.~\ref{fig3040} (right panel).
The convergence to the limit is explicitly guaranteed by properties of the functions $\phi_{k}(\rho|\theta)$.
The limiting forms of these functions as $\rho\to\infty$ are explicitly known for $\theta\in[0,\pi/2]$:
\begin{equation*}
\label{e250}
\phi_{k}(\rho|\theta)\underset{\rho \to \infty}{\sim}\frac{2}{\sqrt{\pi}} \sin{2k\theta}.
\end{equation*}
Therefore, in contrast to the prelimiting case, the smooth behavior of the breakup amplitude near 90 degrees is observed, see the mentioned figures.
%
%
%
%
%
%

\begin{figure}[t]%
\centerline{\includegraphics[width=0.8\textwidth, angle=0.0]{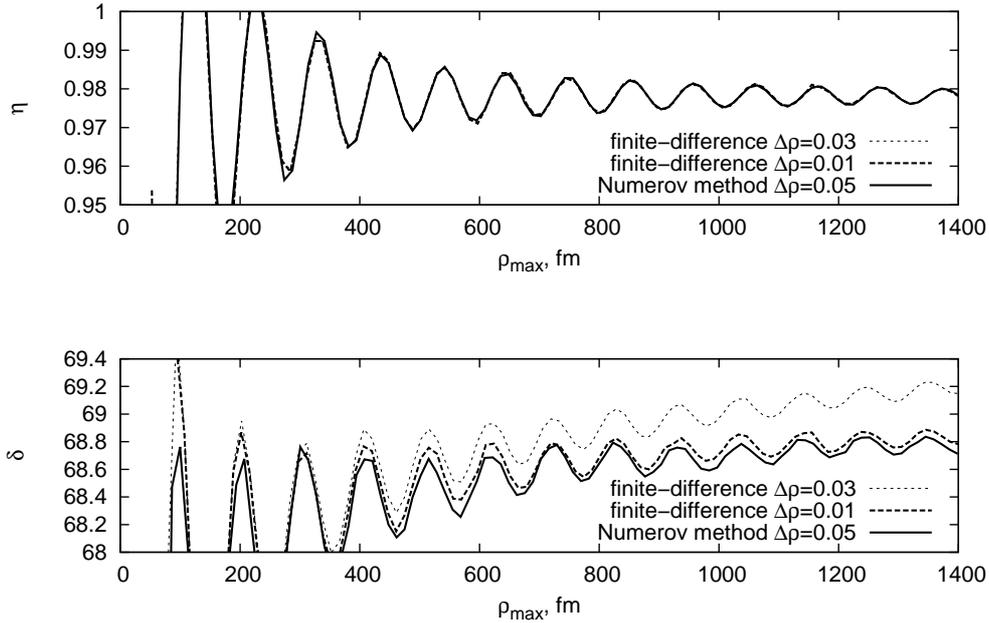}}
\caption{The calculated values of the inelasticity coefficient~(\ref{e260}) $\eta$ (top panel),
phase shift $\delta$ (bottom panel) for $E_{lab}=14.1$~MeV as functions of $\rho_{max}$.
The dashed lines represent the values obtained using the second-order finite-differences~(\ref{e20100}) at the $\rho$-grid with mesh $\Delta\rho=0.03$ and $\Delta\rho=0.01$,
whereas the solid line shows the values obtained using the Numerov method~\cite{Suslov} with $\Delta\rho=0.05$.
The obtained inelasticity coefficients coincide with a good precision.
}
\label{fig50}
\end{figure}

\begin{figure}[t]
\centerline{\includegraphics[width=0.8\textwidth, angle=0.0]{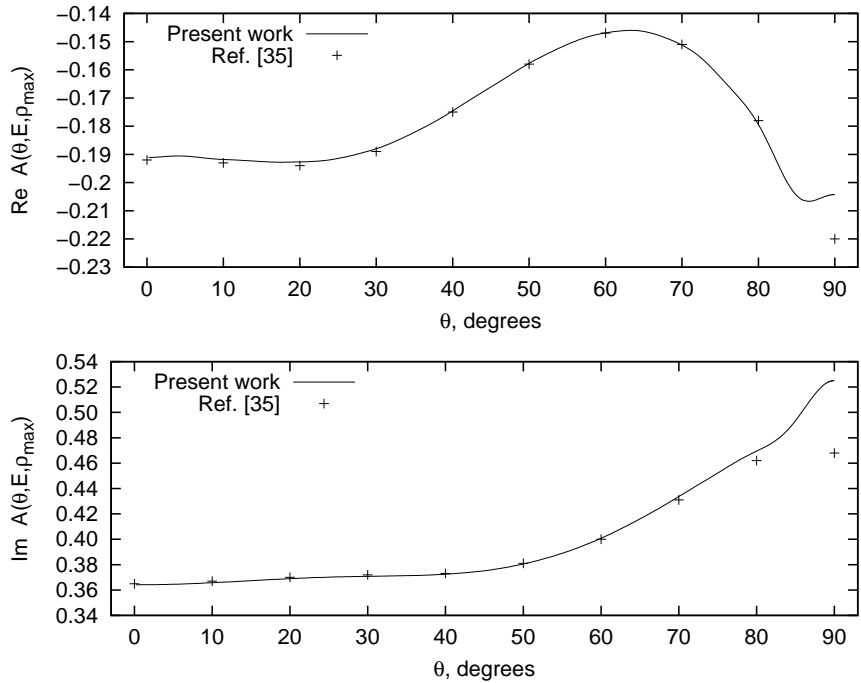}}
\caption{Real and imaginary parts of the breakup amplitude $A(\theta,E,\rho_{max})$ for $E_{lab}=14.1$~MeV.
The results of Ref.~\cite{Friar1995} are also shown for comparison.}
\label{fig60}
\end{figure}

The convergence of the binary amplitude $a_{0}(q,\rho_{max})$ and breakup amplitude $A(\theta,E,\rho_{max})$ as $\rho_{max}\to\infty$
has been obtained.
For example, the $\rho_{max}$-dependence of the inelasticity coefficient, $\eta$, and the phase shift, $\delta$,
defined as
\begin{equation}
\label{e260}
a_{0}=\frac{\eta e^{2i\delta}-1}{2i},
\end{equation}
are presented in Fig.~\ref{fig50}.
The top panel shows that the calculated values of the inelasticity are identical for the three different precisions of the computations.
The bottom panel shows
that the decrease of the mesh step $\Delta\rho$ for the $\rho$-grid to $0.01$~fm leads to
oscillating but significantly less biased values of the phase shift as $\rho_{max}$ increases.
The calculations using the Numerov method~\cite{Suslov} confirm the finite-difference results.
The oscillations are vanishing as $\rho_{max}\to\infty$ and the limiting value of the amplitude can be obtained by extrapolation.
Nevertheless, in order to reach relatively small oscillations it is necessary to achieve values of $\rho_{max}>1000$~fm.
The obtained values of the binary amplitude for different laboratory frame energies
are in a good agreement with the binary amplitudes in Ref.~\cite{Friar1995}.
The obtained breakup amplitude $A(\theta,E,\rho_{max})$ was compared with the results in Ref.~\cite{Friar1995}.
This comparison for $E_{lab}=14.1$~MeV is presented in Fig.~\ref{fig60}.
The results are in good agreement for values of $\theta<80^{\circ}$, whereas
for $\theta\sim90^{\circ}$ the differences are observed.

The typical overall time needed for solution
includes the time for construction of the matrix system, for solving it, and for calculating the scattering amplitudes.
In practical calculations, the first time was comparable with the second one.
The asymptotic approach requires only the last row of the solution of the matrix system,
so the amplitudes are calculated relatively fast.
For example, the overall time of the solution for the number of blocks on the diagonal $N=3000$ with the TA is about 11 hours.
The application of the DM with $P=64$ computing units allowed us to reduce the solution time from 11 hours to about one hour,
i.e. the overall time reduced by a factor of 10.
For larger $N$ and for the method using the integral representation, the overall speedup was not so large, but comparable.
The obtained growth of the performance by a factor of up to ten was practically achieved for various computational studies of the given physical problem.
%
%

\section{Conclusions}
The DM for parallel solution of the block-tridiagonal matrix systems has been described in detail and
the features of the method increasing the performance have been given.
We have shown that rearranging the initial matrix system into equivalent one with the ``arrowhead'' structure of the new matrix
allows one its parallel solving.
The new matrix consists of diagonal independent blocks, sparse auxiliary ones along the right and bottom sides, and the coupling supplementary block of smaller size.
The analytical estimation of the number of multiplicative operations of the method and computational speedup with respect to the serial TA has been performed.
We studied the standard nonrecursive application of the DM as well as the recursive one.
In recursive application, the DM is applied to the initial matrix system and, then, to the obtained independent subsystems and the supplementary matrix equation.
The various cases arised in practice have been considered.
The maximum computational speedup and the parameters providing the achieved speedup have been analytically obtained.
The recursive application of the method allowed us to considerably increase the speedup in comparison to the standard nonrecursive use of the DM.
For the considered cases, the achieved analytical speedup with respect to the TA was varied by a factor up to fifty.
The analytical estimations for the standard nonrecursive case have been validated in practical calculations of solving the matrix system originated from the BVP for the two-dimensional integro-differential Faddeev equations.
The numerical scheme for solution of these equations as well as the illustrative results have been presented.
The overall growth of the performance by a factor of up to ten has been practically achieved in the computations.

\section*{Acknowledgments}
The work was supported by Russian Foundation for Basic Research grant No.~14-02-00326 and by Saint Petersburg State University within
the project No.~11.38.241.2015. All calculations were carried out using the facilities of the ``Computational Center of SPbSU''.


\end{document}